\documentclass[aps,prd,eqsecnum,preprint,tightenlines,nofootinbib,showpacs]{revtex4-1}
\usepackage{graphicx,amsmath,latexsym}

\def\bea{\begin{eqnarray}}
\def\eea{\end{eqnarray}}
\def\be{\begin{equation}}
\def\ee{\end{equation}}
\newcommand{\ub}[1]{\underline{#1}}

\begin{document}

\title{A nonperturbative calculation of the electron's
magnetic moment with truncation extended to two photons}

\author{Sophia S. Chabysheva}
\author{John R. Hiller}
\affiliation{Department of Physics \\
University of Minnesota-Duluth \\
Duluth, Minnesota 55812}

\date{\today}

\begin{abstract}
The Pauli--Villars (PV) regularization scheme is applied
to a calculation of the dressed-electron state and
its anomalous magnetic moment in light-front-quantized quantum
electrodynamics (QED) in Feynman gauge.  The regularization is
provided by heavy, negative-metric fields added to the 
Lagrangian.  The light-front QED Hamiltonian then leads to
a well-defined eigenvalue problem for the dressed-electron state
expressed as a Fock-state expansion.  The Fock-state wave functions
satisfy coupled integral equations that come from this
eigenproblem.  A finite system of equations is obtained by
truncation to no more than two photons and no positrons; this
extends earlier work that was limited to dressing by a single
photon.  Numerical techniques are applied to solve the 
coupled system and compute the anomalous moment, for which we
obtain agreement with experiment, within numerical errors,
but observe a small systematic discrepancy that should be
due to the absence of electron-positron loops and of three-photon
self-energy effects.
We also discuss the prospects for application of the method to
quantum chromodynamics.
\end{abstract}
%%%%%%%%%%%%%%%%%%%%%%%%%%%%%%%%%%%%%%%%%%%%%%%%%%%%%
% 10. THE PHYSICS OF ELEMENTARY PARTICLES AND FIELDS
% 11.10.-z Field theory (for gauge field theories, see 11.15)
% 11.10.Ef Lagrangian and Hamiltonian approach
% 11.10.Gh Renormalization
% 11.15.-q Gauge field theories
% 11.15.Tk Other nonperturbative techniques
% 12. Specific theories and interaction models; particle systematics
% 12.38.-t Quantum chromodynamics see also 24.85 Quarks, gluons,
% and QCD in nuclei and nuclear processes
% 12.38.Lg Other nonperturbative calculations
%%%%%%%%%%%%%%%%%%%%%%%%%%%%%%%%%%%%%%%%%%%%%%%%%%%%%%%

%
\pacs{12.38.Lg, 11.15.Tk, 11.10.Gh, 11.10.Ef
%
%\begin{center}(Submitted to Physical Review D.)\end{center}}
}

\maketitle

%%%%%%%%%%%%%%%%%%%%%%
\section{Introduction}
\label{sec:Introduction}
%%%%%%%%%%%%%%%%%%%%%%

\subsection{Motivation}

High-energy scattering experiments have shown conclusively that
the strong nuclear force is well described by
quantum chromodynamics (QCD), with scattering observables
computed perturbatively.
At longer distance scales, where the properties of hadrons
themselves are determined, the effective couplings are strong 
and nonlinear, making the derivation of hadronic properties
from QCD a difficult task.

Nonperturbative calculations
are always difficult, but for a strongly coupled theory
such as QCD, they are worse.  For a weakly coupled theory,
one can set aside much of the interaction for 
perturbative treatment and solve only a small core
problem nonperturbatively.  For QED, this core problem
is the Coulomb problem; when
combined with high-order perturbation theory, amazingly
accurate results can be obtained for bound states of
the theory~\cite{Kinoshita}.  In a strongly coupled theory one cannot
make this separation so easily.

In the work presented here, the purpose
is to explore a nonperturbative method that
can be used to solve for the bound states of quantum field theories.
Although the bound states of QCD are of particular interest, the
method is not yet mature enough for application to QCD.
Instead, we will continue with the program developed in the
earlier work of Brodsky, McCartor, and 
Hiller~\cite{bhm1,bhm2,YukawaDLCQ,ExactSolns,YukawaOneBoson,OnePhotonQED,YukawaTwoBoson},
and more recently continued by us~\cite{ChiralLimit,thesis,SecDep},
and explore the method within QED.
This provides an analysis of a gauge theory,
which is a critical step toward solving a non-Abelian gauge theory,
such as QCD.

\subsection{Fock-state expansions and Pauli--Villars regularization}

We will use Fock states
as the basis for the expansion of eigenstates.  
Each bound state is an eigenstate of
the field-theoretic Hamiltonian, and projections of this
eigenproblem onto individual Fock states yields coupled
equations for the Fock-state wave functions.
We truncate the expansion to have a calculation of
finite size.

The solution of such equations, in general, requires numerical
techniques.  The equations are converted to a matrix 
eigenvalue problem by some discretization of the 
integrals~\cite{PauliBrodsky}
or by a function expansion for the wave functions~\cite{Varyetal}.  
The matrix is usually large
and not diagonalizable by standard techniques; instead, one
or some of the eigenvalues and eigenvectors are extracted by 
the iterative Lanczos process~\cite{Lanczos,Cullum}.
The eigenvector of the matrix yields the wave functions, and from
these can be calculated the properties of the eigenstate, by
considering expectation values of physical observables.

Although this may seem straightforward, 
the integrals of the integral equations are not finite
and must be regulated in some way.  We use
Pauli--Villars (PV) regularization~\cite{PauliVillars}
for these ultraviolet divergences.
The basic idea is to subtract from each integral a
contribution of the same form but of a PV particle
with a much larger mass. This subtraction will cancel
the leading large-momentum behavior of the integrand,
making the integral less singular.
To make an integral
finite, more than one subtraction
may be necessary, due to subleading divergences.
To arrange the subtractions, we assign the PV particles a
negative metric. 
The masses of these PV particles are then
the regulators of the redefined theory, and ideally one
would take the limit of infinite PV masses at the end of
the calculation.

Ordinarily, this method of regularization, being 
automatically relativistically covariant, preserves
the original symmetries of the theory.  However,
it may happen that the negative-metric PV particles
over-subtract, in the sense that some symmetry is
broken by a finite amount.  In such a case, a
counterterm is needed, or a positive-metric PV
particle can be added to restore the symmetry.
We have shown in \cite{ChiralLimit} that
a positive-metric PV photon does restore
the correct chiral limit.

It is interesting to note that
the introduction of negative-metric partners has recently
been used to define extensions of the Standard Model
that solve the hierarchy problem~\cite{Lebed}.  The
additional fields provide cancellations that reduce
the ultraviolet divergence of the bare Higgs mass
to only logarithmic.  This slowly varying dependence
allows the remaining cancellations to occur
without excessive fine tuning.

A serious complication in the use of
Fock-state expansions and coupled equations
is the presence of vacuum contributions to
the eigenstate.  The lack of particle-number
conservation in quantum field theory means that,
in general, even the vacuum can have contributions
from nonempty Fock states with zero momentum
and zero charge.  The basis for a massive eigenstate
will include such vacuum Fock states in products 
with nonvacuum Fock states, since the vacuum
contributions do not change the momentum or
charge.  These vacuum contributions destroy the
interpretation of the wave functions.
In order to have well-defined Fock-state
expansions and a simple vacuum, we use the light-cone
coordinates of Dirac~\cite{Dirac,DLCQreview}.
Light-cone coordinates also have the advantage of separating
the internal and external momenta of a system.  The Fock-state
wave functions depend only on the internal momenta.  The state
can then be boosted to any frame without necessitating the
recalculation of the wave functions.

\subsection{Numerical methods and Fock-space truncation}

The standard approach to numerical solution of the eigenvalue problem
is the method originally suggested by Pauli and
Brodsky~\cite{PauliBrodsky}, discrete light-cone quantization (DLCQ).
Periodic boundary conditions are applied in a light-cone box
of finite size, and the light-cone momenta are resolved to a
discrete grid.  Because this method can be formulated at the
second-quantized level, it provides for the systematic inclusion
of higher Fock sectors.  DLCQ has been particularly successful
for two-dimensional theories, including QCD~\cite{Hornbostel}
and supersymmetric Yang--Mills theory~\cite{SDLCQ}.  There was
also a very early attempt by Hollenberg {\em et al}.~\cite{EarlyLCQCD}
to solve four-dimensional QCD.

Unfortunately, the kernels of the QED
integral operators require a very fine DLCQ grid if the 
contributions from heavy PV particles are to be accurately
represented.  To keep the discrete matrix eigenvalue problem
small enough, we use instead the
discretization developed for the analogous problem in Yukawa 
theory~\cite{YukawaTwoBoson}, suitably adjusted for the 
singularities encountered in QED.

An explicit truncation in particle number, the light-cone equivalent
of the Tamm--Dancoff approximation~\cite{TammDancoff}, can
be made.  This truncation has significant consequences
for the renormalization of the theory~\cite{SectorDependent,Wilson},
in particular the uncancelled divergences discussed below.  It
also impacts comparisons to Feynman perturbation 
theory~\cite{RecentPert},
where the truncation eliminates some of the time-ordered graphs
that are required to construct a complete Feynman graph.  Fortunately,
numerical tests in Yukawa theory~\cite{YukawaDLCQ,YukawaTwoBoson}
indicate that these difficulties can be overcome.  The tests show
a rapid convergence with respect to particle number.

To carry out our calculation in QED, three problems must be 
solved, as discussed in \cite{OnePhotonQED}.  
We need to respect gauge invariance, interpret new singularities from energy
denominators, and handle uncancelled divergences.  Although PV regularization
normally preserves gauge invariance, the flavor-changing interactions
chosen for the PV couplings, where emission or absorption of a photon can
change the flavor of the fermion, do break the invariance at finite mass values
for the PV fields;
we assume that an exact solution exists and has all symmetries
and that a close approximation can safely break symmetries.  The new
singularities occur because the bare mass of the electron is less
than the physical mass and energy denominators can be zero; a
principal-value prescription is used.  These zeros have the appearance
of a threshold but do not correspond to any available decay.   The
uncancelled divergences are handled (as in the case of 
Yukawa theory~\cite{YukawaTwoBoson}),
with the PV masses kept finite and the finite-PV-mass error balanced
against the truncation error.

For small PV masses, too much of the negatively normed states are
included in the eigenstate.  For large PV masses, there are truncation
errors; the exact eigenstate has large projections onto excluded Fock 
sectors.  To see the effect of truncation, consider the form of the
coupling dependence for the anomalous magnetic moment, which is
\be
\frac{a_1 g^2 \,[+a_2 g^4 \ln\mu_{\rm PV}+\cdots]}
                  {1+b_1g^2+b_2g^2\ln\mu_{\rm PV}+\cdots},
\ee
where $\mu_{\rm PV}$ is a PV mass scale.  The contents of
the square brackets are absent in the case of truncation.
When the large-$\mu_{\rm PV}$ limit is taken, this expression 
becomes zero when there is truncation and a nonzero, finite number
when there is not.  In perturbation theory, the order-$g^2$
terms in the denominator are kept only if the order-$g^4$ terms
are kept in the numerator, and a finite result is again obtained
in the large-$\mu_{\rm PV}$ limit.  However, the truncated
nonperturbative calculation includes the order-$g^2$ terms in
the denominator but not the compensating order-$g^4$ terms in 
the numerator.  The associated error is minimized by keeping
$\mu_{\rm PV}$ as small as possible, but if too small, the
errors associated with keeping the unphysical PV Fock states
in the basis will be too large.  A compromise is to be found
for a range of intermediate values of $\mu_{\rm PV}$ for which
physical quantities are independent of $\mu_{\rm PV}$.
For QED, we see this in the behavior of the anomalous
moment of the electron, as a function of the PV masses.

For Yukawa theory, the usefulness of truncating Fock space
was checked in a DLCQ calculation that included many Fock sectors.
The full DLCQ result was compared with results for truncations to a few Fock sectors for
weak to moderate coupling strengths and found
to agree quite well~\cite{YukawaDLCQ}.  
We can see in Table 1 of \cite{YukawaDLCQ} that probabilities for 
higher Fock states decrease rapidly.  This was also
checked at stronger coupling by comparing the two-boson and one-boson 
truncations~\cite{YukawaTwoBoson}. Figure 14 of \cite{YukawaTwoBoson}
shows that contributions to structure
functions from the three-particle sector are much smaller than those
from the two-particle sector.

For QED, there has been no explicit demonstration that truncation in
Fock space is a good approximation; the two-photon truncation
considered here gives the first evidence.  The usefulness
of truncation is expected for general
reasons, but a physical argument comes from comparing perturbation
theory with the Fock-space expansion.  Low-order truncations in 
particle number correspond to doing perturbation theory in $\alpha$ 
to low order, plus keeping partial contributions for all orders in 
$\alpha$.  As long as the theory is regulated 
so that the contributions are finite, the contributions of higher 
Fock states are expected to be small because they are higher order in $\alpha$. 
Of course, due to limitations on numerical accuracy, we do
not expect to be able to compute the anomalous moment as
accurately as high-order perturbative calculations~\cite{Kinoshita,Langnau}.

\subsection{Applications of the method}

The PV regularization method has been considered for QED
and applied to a one-photon truncation of the dressed
electron state~\cite{OnePhotonQED,ChiralLimit,thesis,SecDep}.
In Feynman gauge,
one PV electron and one PV photon were sufficient if the
PV electron mass is taken to infinity; otherwise, a second
PV photon is needed to restore the chiral limit~\cite{ChiralLimit}.
This choice of regularization has the convenient feature of not only
cancelling the instantaneous fermion interactions but
also making the fermion constraint equation explicitly
solvable.  

The cancellation of the instantaneous fermion interactions occurs
because the individual contributions are independent of the fermion
mass and have opposite signs.  The instantaneous contributions usually
provide important infrared cancellations, making their absence a
possible cause for concern, but these infrared cancellations are instead
provided by PV contributions.  The absence of the instantaneous
terms is important for the numerical calculation, because these terms
can greatly increase the computational load, and is significant
compensation for the addition of the PV fields to the basis.

Ordinarily, in the light-cone quantization 
of QED~\cite{Tang}, light-cone gauge ($A^+=0$)
must be chosen to make the fermion constraint equation 
solvable; in Feynman gauge, with one PV electron and
one or two PV photons, the $A^+$ terms cancel from the constraint
equation.  Light-cone gauge was considered
in \cite{OnePhotonQED}, but the naive choice of
three PV electrons for regularization was found
insufficient; an additional photon and higher
derivative counterterms were also needed.  The
one-photon truncation yielded an anomalous moment
within 14\% of the Schwinger term~\cite{Schwinger}.
For the two-photon truncation considered here,
the value for the 
anomalous moment should be close to the value obtained
perturbatively when the Sommerfield--Petermann
term~\cite{SommerfieldPetermann} is included.
However, numerical errors will make this tiny correction
undetectable, and we will focus on obtaining better
agreement with the leading Schwinger term of $\alpha/2\pi$.

An extension to a two-boson truncation is
interesting as a precursor
to work on QCD.  Unlike the one-boson truncation, where QED and QCD are
effectively indistinguishable, the two-boson truncation allows three and
four-gluon vertices to enter the calculation.  A nonperturbative 
calculation, with these nonlinearities included, could capture
much of the low-energy physics of QCD, perhaps even confinement.

The approach depends critically on making a 
Tamm--Dancoff truncation to a finite number of
constituents.  For QCD this is thought to be reasonable
because the constituent quark model was so 
successful~\cite{CQM}.
Wilson and collaborators~\cite{quarkonia,Wilson} even argued that a 
light-cone Hamiltonian approach can provide an explanation
for the quark model's success.
The recent successes of the AdS/CFT correspondence~\cite{AdSQCD} 
in representing the light hadron spectrum of QCD
also indicates the effectiveness of
a truncation; this description of hadrons is equivalent to
keeping only the lowest valence light-cone Fock state.

At the very least, the success of the constituent quark model
shows that there exists an effective description of the bound
states of QCD in
terms of a few degrees of freedom.  It is likely that
the constituent quarks of the quark model correspond to
effective fields, the quarks of QCD dressed by gluons
and quark-antiquark pairs.
From the exact solutions obtained using PV 
regularization in the unphysical equal-mass limit~\cite{ExactSolns}, 
it is known that simple Fock states in light-cone quantization correspond
to very complicated states in equal-time quantization,
and this structure may aid in providing some correspondence
to the constituent quarks.  However, the truncation of the QCD Fock space
may need to be large enough to include states that 
provide the dressing of the current quarks, and perhaps a
sufficiently relaxed truncation is impractical.  As an alternative,
the light-front PV method could be applied to an effective
QCD Lagrangian in terms of the effective fields.  Some work on
developing a description of light-front QCD in terms of effective
fields has been done by G{\l}azek {\em et al}.~\cite{Glazek}.

\subsection{Other methods}

A directly related light-front Hamiltonian approach is that of sector-dependent
renormalization~\cite{SectorDependent}, where bare masses and couplings
are allowed to depend on the Fock sector.  This alternative treatment
was used by Hiller and Brodsky~\cite{hb} and more recently by 
Karmanov, Mathiot, and Smirnov~\cite{Karmanov}.  In principle, this
approach is roughly equivalent to the approach used here; however,
the authors of \cite{Karmanov} neglect the limitations on the PV masses that
come from having a finite, real bare coupling, as discussed in \cite{hb}
and \cite{SecDep},
and do not make the projections necessary to have finite expectation
values for particle numbers.

Our method is complementary to lattice gauge
theory~\cite{lattice}, which has been studied for much longer than
nonperturbative light-front methods and has already achieved impressive
successes in solving QCD.  However, it is formulated in Euclidean spacetime
and has
particular difficulty with quantities such as timelike and spacelike
form factors, that depend on the signature of the Minkowski metric.
In contrast, in a Hamiltonian approach with the original Minkowski
metric, a form factor is readily calculated as a convolution of
wave functions.

A related method is that of the transverse lattice~\cite{TransLattice},
where light-cone methods are used for the longitudinal direction and
lattice methods for the transverse.  It is, however, a Hamiltonian
approach which results in wave functions. 
Applications have been to large-$N$ gauge
theories and mostly limited to consideration of meson and glueball
structure.

Another approach is that of
Dyson--Schwinger equations~\cite{SchwingerDyson}, which are
coupled equations for the $n$-point Euclidean Green's functions
of a theory, including the propagators for the fundamental
fields.  Bound states of $n$ constituents appear as poles in
the $n$-particle propagator.  Solution
of the infinite system requires truncation and a model for the
highest $n$-point function. 
Again, as in the lattice approach, there is the limitation 
to a Euclidean formulation.

\subsection{Outline}

The content of the remainder of the paper is as follows.  The
structure of the light-front Hamiltonian and the Fock-state
expansion of the eigenstate are presented in Sec.~\ref{sec:dressedelectron},
where these are used to obtain coupled integral equations for
the wave functions.  Expressions for the normalization and
anomalous magnetic moment are also given.  Section~\ref{sec:twophoton}
contains the discussion of the solution of the coupled equations,
with results for the anomalous moment presented in Sec.~\ref{sec:results}.
A summary of the work is given in Sec.~\ref{sec:summary}.  Details
of the numerics are left to Appendices.

\section{THE DRESSED-ELECTRON EIGENSTATE} \label{sec:dressedelectron}

\subsection{Helicity basis}

For calculations with more than one photon in the Fock space,
an helicity basis is convenient.  The dependence of the
vertex functions on azimuthal angle then becomes simple.
This will allow us to take advantage
of cylindrical symmetry in the integral equations, such that
the azimuthal angle dependence can be handled analytically.
There is then no need to discretize the angle in making the
numerical approximation.  To introduce the helicity basis,
we define the following annihilation operators for the photon 
fields\footnote{For details of Feynman-gauge QED on the light 
front, particularly the notation,
see the discussion in \protect\cite{OnePhotonQED} 
and \protect\cite{ChiralLimit}.  For a discussion of the residual
gauge freedom and the projection onto the physical subspace,
see \protect\cite{ChiralLimit}.}

\be \label{eq:helicityops}
a_{l\pm}=\frac{1}{\sqrt{2}}(a_{l0}\pm a_{l3})\,, \;\; 
a_{l(\pm)}=\frac{1}{\sqrt{2}}(a_{l1}\pm ia_{l2}) .
\ee
The Hamiltonian can then be rearranged to the form
\bea \label{eq:HelicityP-}
\lefteqn{{\cal P}^-=
   \sum_{i,s}\int d\ub{p}
      \frac{m_i^2+p_\perp^2}{p^+}(-1)^i
          b_{i,s}^\dagger(\ub{p}) b_{i,s}(\ub{p})} \\
   && +\sum_{l,\lambda}\int d\ub{k}
          \frac{\mu_l^2+k_\perp^2}{k^+}(-1)^l
         \left[-a_{l\lambda}^\dagger(\ub{k}) a_{l,-\lambda}(\ub{k}) 
               +a_{l(\lambda)}^\dagger(\ub{k}) a_{l(\lambda)}(\ub{k})
                                           \right]
          \nonumber \\
   && +\sum_{i,j,l,s,\lambda}\int d\ub{p} d\ub{q}\sqrt{\xi_l}\left\{
      b_{i,s}^\dagger(\ub{p}) b_{j,s}(\ub{q})
      \left[ V_{ij,2s}^\lambda(\ub{p},\ub{q})a_{l\lambda}^\dagger(\ub{q}-\ub{p}) 
      \right. \right.      \nonumber \\
   && \rule{2.5in}{0in} \left.  
          +V_{ij,2s}^{(\lambda)}(\ub{p},\ub{q})a_{l(\lambda)}^\dagger(\ub{q}-\ub{p})
	                      \right]   \nonumber \\
      &&\left.\rule{0.5in}{0in}       
+b_{i,s}^\dagger(\ub{p}) b_{j,-s}(\ub{q})
      \left[ U_{ij,-2s}^\lambda(\ub{p},\ub{q})a_{l\lambda}^\dagger(\ub{q}-\ub{p})
             +U_{ij,-2s}^{(\lambda)} a_{l(\lambda)}^\dagger(\ub{q}-\ub{p})\right]
                    + H.c.\right\}\,,  \nonumber
\eea
and the vertex functions become
\bea \label{eq:HelicityVertices}
V_{ij\pm}^+(\ub{p},\ub{q})&=&\frac{e}{\sqrt{8\pi^3(q^+-p^+)}} , \\
V_{ij\pm}^-(\ub{p},\ub{q})&=&\frac{e}{\sqrt{8\pi^3}}
             \frac{(p^1\mp i p^2)(q^1\pm i q^2)+m_i m_j}
                   {p^+q^+\sqrt{q^+-p^+}} , \nonumber\\
V_{ij\pm}^{(\pm)}(\ub{p},\ub{q})&=&\frac{e}{\sqrt{8\pi^3}}
             \frac{q^1\pm iq^2}{q^+\sqrt{q^+-p^+}} ,\nonumber \\
V_{ij\mp}^{(\pm)}(\ub{p},\ub{q})&=&\frac{e}{\sqrt{8\pi^3}}
             \frac{p^1\pm ip^2}{p^+\sqrt{q^+-p^+}} ,  \nonumber\\
U_{ij\pm}^+(\ub{p},\ub{q})&=&0 , \nonumber \\
U_{ij\pm}^-(\ub{p},\ub{q})&=&\mp\frac{e}{\sqrt{8\pi^3}}
             \frac{m_j(p^1\pm ip^2)-m_i(q^1\pm iq^2)}
                                {p^+q^+\sqrt{q^+-p^+}} ,  \nonumber\\
U_{ij\pm}^{(\pm)}(\ub{p},\ub{q})&=&0 , \nonumber \\
U_{ij\mp}^{(\pm)}(\ub{p},\ub{q})&=&\mp\frac{e}{\sqrt{8\pi^3}}
             \frac{m_i q^+-m_j p^+}{p^+q^+\sqrt{q^+-p^+}} . \nonumber
\eea
The $a_{l\pm}$ operators are null, in the sense that 
$[a_{l\pm}(\ub{k}),a_{l'\pm}^\dagger(\ub{k}')]=0$;
however, we do have 
$[a_{l\pm}(\ub{k}),a_{l'\mp}^\dagger(\ub{k}')]=-\delta_{ll'}\delta(\ub{k}-\ub{k}')$.

We will study the state of the electron as an eigenstate of this 
light-cone Hamiltonian.  In general, the electron
is dressed by photons and electron-positron
pairs; however, we limit the calculation to photons and truncate
the number of photons to two, at most.  The eigenstate is then expanded in terms
of Fock states.  In order that the Fock expansion
be an eigenstate of the light-cone Hamiltonian, the Fock-state
wave functions must
satisfy coupled integral equations.  The wave functions are also
constrained by normalization of the state.  The anomalous magnetic
moment is then calculated from a spin-flip matrix element.
In the remainder of this section, we collect the 
fundamental expressions for the Fock-state expansion, the coupled
equations for the wave functions, the normalization of the wave functions,
and the anomalous moment.

\subsection{Fock-state expansion} \label{sec:FockStateExpansion}
   
It is convenient to work in a Fock basis where ${\cal P}^+$
and $\vec{\cal P}_\perp$ are diagonal and the total transverse
momentum $\vec P_\perp$ is zero.  
We expand the eigenfunction for the dressed-fermion state
with total $J_z=\pm \frac12$ in such a Fock basis as
\bea \label{eq:FockExpansion}
\lefteqn{|\psi^\pm(\ub{P})\rangle=\sum_i z_i b_{i\pm}^\dagger(\ub{P})|0\rangle
  +\sum_{ijs\mu}\int d\ub{k} C_{ijs}^{\mu\pm}(\ub{k})b_{is}^\dagger(\ub{P}-\ub{k})
                                       a_{j\mu}^\dagger(\ub{k})|0\rangle}&& \\
 && +\sum_{ijks\mu\nu}\int d\ub{k_1} d\ub{k_2} C_{ijks}^{\mu\nu\pm}(\ub{k_1},\ub{k_2})
       \frac{1}{\sqrt{1+\delta_{jk}\delta_{\mu\nu}}}  
              b_{is}^\dagger(\ub{P}-\ub{k_1}-\ub{k_2})
                 a_{j\mu}^\dagger(\ub{k_1})a_{k\nu}^\dagger(\ub{k_2})|0\rangle, 
     \nonumber
\eea
where we have truncated the expansion to include at most two photons.
The $z_i$ are the amplitudes for the bare electron states, with $i=0$ for
the physical electron and $i=1$ for the PV electron.  The $C_{ijs}^{\mu\pm}$ are
the two-body wave functions for Fock states with an electron of flavor $i$
and spin component $s$ and a photon of flavor $j=0$, 1 or 2 and field component $\mu$,
expressed as functions of the photon momentum.  The upper index of $\pm$
refers to the $J_z$ value of $\pm\frac12$ for the eigenstate.
Similarly, the $C_{ijks}^{\mu\nu\pm}$ are the three-body wave functions
for the states with one electron and two photons, with flavors $j$ and $k$
and field components $\mu$ and $\nu$.

Careful interpretation of the eigenstate is required to
obtain physically meaningful answers.  In particular,
there needs to be a physical state with positive norm.
We apply the same approach as was used in Yukawa
theory~\cite{YukawaOneBoson}.
A projection onto the physical subspace is accomplished
by expressing Fock states in terms of positively normed
creation operators $a_{0\mu}^\dagger$, $a_{2\mu}^\dagger$, and 
$b_{0s}^\dagger$ and the null combinations 
$a_\mu^\dagger=\sum_i \sqrt{\xi_i}a_{i\mu}^\dagger$ 
and $b_s^\dagger=b_{0s}^\dagger+b_{1s}^\dagger$.
The $b_s^\dagger$ particles are annihilated by the
generalized electromagnetic current $\bar\psi\gamma^\mu\psi$; thus,
$b_s^\dagger$ creates unphysical contributions to be dropped,
and, by analogy, we also drop contributions created by
$a_\mu^\dagger$.  The projected dressed-fermion state is
\bea \label{eq:projected}
|\psi^\pm(\ub{P})\rangle_{\rm phys}&=&\sum_i (-1)^i z_i
                                          b_{0\pm}^\dagger(\ub{P})|0\rangle \\
  &&+\sum_{s\mu}\int d\ub{k} \sum_{i=0}^1\sum_{j=0,2}\sqrt{\xi_j}
        \sum_{k=j/2}^{j/2+1} \frac{(-1)^{i+k}}{\sqrt{\xi_k}}
                        C_{iks}^{\mu\pm}(\ub{k})
               b_{0s}^\dagger(\ub{P}-\ub{k})
                                       a_{j\mu}^\dagger(\ub{k})|0\rangle \nonumber \\
 && +\sum_{s\mu\nu}\int d\ub{k_1} d\ub{k_2} 
            \sum_{i=0}^1 \sum_{j,k=0,2} \sqrt{\xi_j\xi_k}
                \sum_{l=j/2}^{j/2+1} \sum_{m=k/2}^{k/2+1} 
                  \frac{(-1)^{i+l+m}}{\sqrt{\xi_l\xi_m}}
              \frac{C_{ilms}^{\mu\nu\pm}(\ub{k_1},\ub{k_2})}
                          {\sqrt{1+\delta_{lm}\delta_{\mu\nu}}}
            \nonumber \\
  &&  \rule{1in}{0mm} \times    b_{0s}^\dagger(\ub{P}-\ub{k_1}-\ub{k_2})
                 a_{j\mu}^\dagger(\ub{k_1})a_{k\nu}^\dagger(\ub{k_2})|0\rangle .
     \nonumber
\eea
This projection is to be used to compute the anomalous moment.

Before using these states, it is important to consider
the renormalization of the external coupling to the 
charge~\cite{BRS,ChiralLimit}.
We exclude fermion-antifermion states, and, therefore,
there is no vacuum polarization.  Thus, if the vertex and
wave function renormalizations cancel, there will be no
renormalization of the external coupling.  As shown in
\cite{ChiralLimit},
this is what happens, but only for the plus component of the current.
Our calculations of the anomalous moment are therefore based on matrix elements of 
the plus component and do not require additional renormalization.

\subsection{Coupled integral equations}

\begin{figure}[ht]
\vspace{0.1in}
\centerline{\includegraphics[width=12cm]{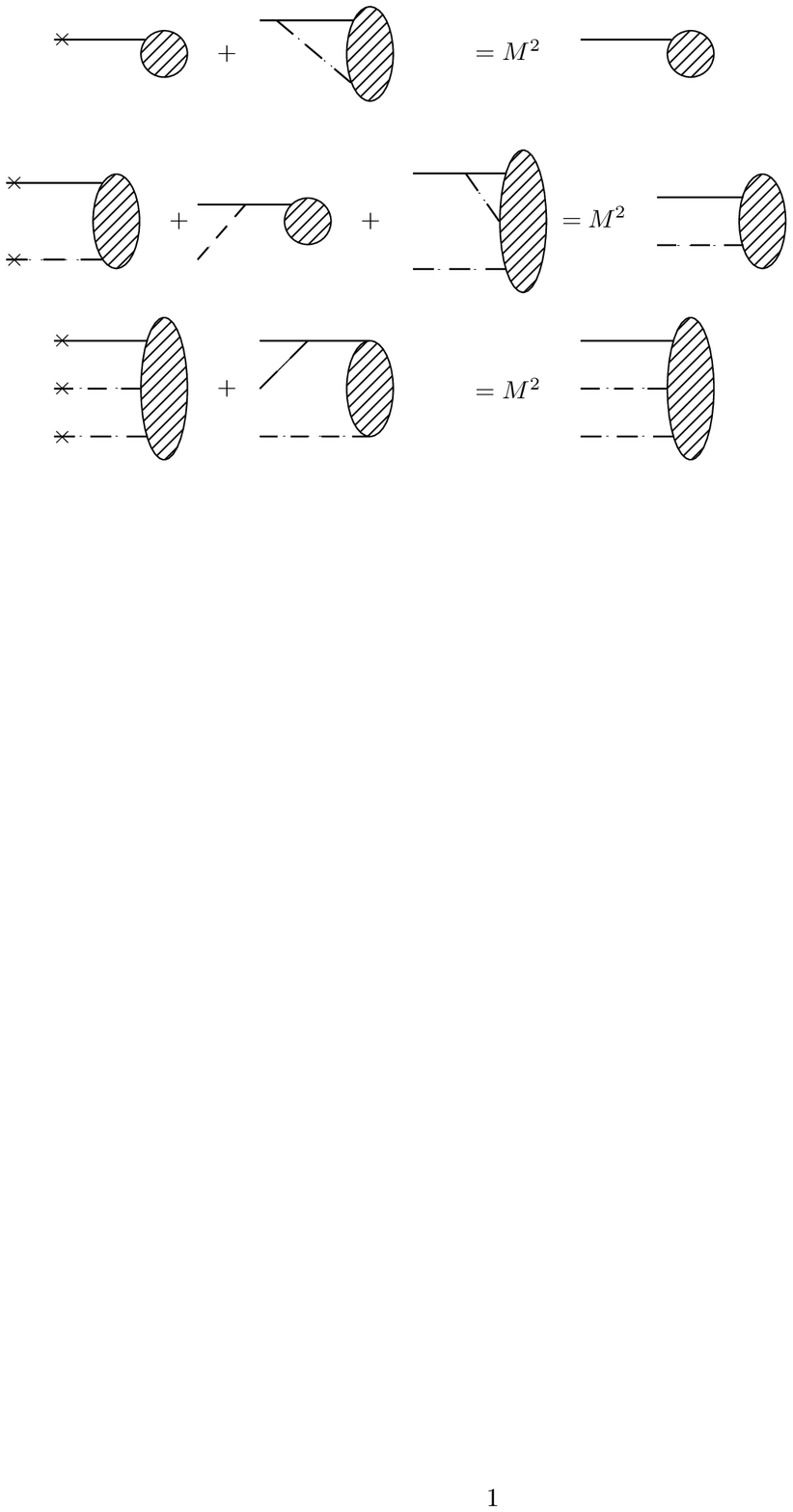}}
\caption[Diagrammatic representation of
the coupled equations.]%
{\label{fig:viscacha1} Diagrammatic representation of
the coupled equations (\ref{eq:firstcoupledequation}),
(\ref{eq:secondcoupledequation}), and 
(\ref{eq:thirdcoupledequation}) of the text.  The filled circles and ovals
represent wave functions for Fock states; the solid lines represent
fermions; and the dashed lines represent photons.  The crosses on
lines represent the light-cone kinetic energy contributions, which
are summed over all particles in the Fock state.}
\end{figure}

The bare amplitudes $z_i$ and
wave functions $C_{ijs}^{\mu\pm}$ and $C_{ijks}^{\mu\nu\pm}$
that define the eigenstate must satisfy the coupled 
system of equations that results from the field-theoretic
mass-squared eigenvalue problem
\begin{equation} \label{eq:EigenProb}
P^-|\psi^\pm(\ub{P})\rangle=\frac{M^2}{P^+}|\psi^\pm(\ub{P})\rangle .
\end{equation}
We work in a frame where the total transverse momentum
is zero and require that this state be an eigenstate
of $P^-$ with eigenvalue $M^2/P^+$.  The form of $P^-$
is given in Eq.~(\ref{eq:HelicityP-}).  
The wave functions then satisfy the following
coupled integral equations:
\bea \label{eq:firstcoupledequation}
[M^2-m_i^2]z_i & = & \int d\ub{q} \sum_{j,l,\mu}\sqrt{\xi_l}(-1)^{j+l}\epsilon^\mu P^+
  \left[V_{ji\pm}^{\mu*}(\ub{P}-\ub{q},\ub{P})C_{jl\pm}^{\mu\pm}(\ub{q}) \right. \\
   &&  \rule{1in}{0mm} \left.   
     +U_{ji\pm}^{\mu*}(\ub{P}-\ub{q},\ub{P}) C_{jl\mp}^{\mu\pm}(\ub{q})\right], \nonumber
\eea
\bea \label{eq:secondcoupledequation}
\lefteqn{\left[M^2 - \frac{m_i^2 + q_\perp^2}{(1-y)} - \frac{\mu_l^2 + q_\perp^2}{y}\right]
  C_{ils}^{\mu\pm}(\ub{q}) }&& \\
    &=& \sqrt{\xi_l}\sum_j (-1)^j z_j P^+ 
      \left[\delta_{s,\pm 1/2}V_{ijs}^\mu(\ub{P}-\ub{q},\ub{P})
             +\delta_{s,\mp 1/2}U_{ij,-s}^\mu(\ub{P}-\ub{q},\ub{P})\right] \nonumber \\
    && +\sum_{ab\nu}(-1)^{a+b}\epsilon^\nu\int d\ub{q}' 
          \frac{2\sqrt{\xi_b}}{\sqrt{1+\delta_{bl}\delta^{\mu\nu}}}
    \left[V_{ais}^{\nu *}(\ub{P}-\ub{q}'-\ub{q},\ub{P}-\ub{q}')
             C_{abls}^{\nu\mu\pm}(\ub{q}',\ub{q}) \right. \nonumber \\
    &&   \rule{2in}{0mm} \left.  +U_{ais}^{\nu *}(\ub{P}-\ub{q}'-\ub{q},\ub{P}-\ub{q}')
             C_{abl,-s}^{\nu\mu\pm}(\ub{q}',\ub{q}) \right] , \nonumber
\eea
\bea \label{eq:thirdcoupledequation}
\lefteqn{\left[M^2 - \frac{m_i^2 + (\vec q_{1\perp}+\vec q_{2\perp})^2}{(1-y_1-y_2)} 
      - \frac{\mu_j^2 + q_{1\perp}^2}{y_1}  - \frac{\mu_l^2 + q_{2\perp}^2}{y_2}\right]
  C_{ijls}^{\mu\nu\pm}(\ub{q}_1,\ub{q}_2) }&& \\
    &=& \frac{\sqrt{1+\delta_{jl}\delta^{\mu\nu}}}{2}\sum_a (-1)^a \left\{
        \sqrt{\xi_j}\left[V_{ias}^\mu(\ub{P}-\ub{q}_1-\ub{q}_2,\ub{P}-\ub{q}_2)
                              C_{als}^{\nu\pm}(\ub{q}_2)  \right.  \right.  \nonumber \\
     && \rule{2in}{0mm}  \left.   +U_{ia,-s}^\mu(\ub{P}-\ub{q}_1-\ub{q}_2,\ub{P}-\ub{q}_2)
                              C_{al,-s}^{\nu\pm}(\ub{q}_2)\right]  \nonumber \\
      &&\rule{1in}{0mm}  
        + \sqrt{\xi_l}\left[V_{ias}^\nu(\ub{P}-\ub{q}_1-\ub{q}_2,\ub{P}-\ub{q}_1)
                              C_{ajs}^{\mu\pm}(\ub{q}_1)   \right. \nonumber \\
     &&  \left.\left. \rule{2in}{0mm}
                   +U_{ia,-s}^\nu(\ub{P}-\ub{q}_1-\ub{q}_2,\ub{P}-\ub{q}_1)
                              C_{aj,-s}^{\mu\pm}(\ub{q}_1)\right] \right\}. \nonumber
\eea
A diagrammatic representation is given in Fig.~\ref{fig:viscacha1}.%
%%%%%%%%%%%%%%%%%%%%%%%%%%%%%%%%%%%%%
%%%%%%%%%%%%%%%%%%%%%%%%%%%%%%%%%%%%%%%%%%%%%%%%%%%%%%%%%%%%%%
The first of these equations 
couples the bare amplitudes $z_i$ to the 
two-body wave functions, $C_{ijs}^{\mu\pm}$.  The second couples the
$C_{ijs}^{\mu\pm}$ to the $z_i$ and to the three-body wave functions,
$C_{ijks}^{\mu\nu\pm}$.  The third is truncated, with no four-body terms,
and simply couples $C_{ijks}^{\mu\nu\pm}$ to $C_{ijs}^{\mu\pm}$
algebraically.
From the structure of the equations, one can show that
the two-body wave functions for the $J_z=-1/2$ eigenstate are 
related to the $J_z=+1/2$ wave functions by
\be
C_{ij+}^{\mu-}=-C_{ij-}^{\mu+*},\;\;
C_{ij-}^{\mu-}=C_{ij+}^{\mu+*}.
\ee
This will be useful in computing the spin-flip matrix element needed
for the anomalous moment.

\subsection{Normalization and anomalous moment}

The projected Fock expansion (\ref{eq:projected}) is normalized
according to
\be  \label{eq:norm}
\langle\psi^{\sigma'}(\ub{P}')|\psi^\sigma(\ub{P})\rangle_{\rm phys}
                  =\delta(\ub{P}'-\ub{P})\delta_{\sigma'\sigma}.
\ee
In terms of the wave functions, this becomes                  
\bea  \label{eq:TwoPhotonNorm}
1&=&|\sum_i(-1)^iz_i|^2
+\sum_{s\mu}\int d\ub{k}\epsilon^\mu \sum_{j=0,2}\xi_j \left|
  \sum_{i=0}^1\sum_{k=j/2}^{j/2+1}
    \frac{(-1)^{i+k}}{\sqrt{\xi_k}}C_{iks}^{\mu+}(\ub{k})\right|^2 \\
&& +\sum_{s\mu\nu}\int d\ub{k_1} d\ub{k_2} \sum_{j,k=0,2} \xi_j\xi_k
            \left| \sum_{i=0}^1 
                \sum_{l=j/2}^{j/2+1} \sum_{m=k/2}^{k/2+1} 
                  \frac{(-1)^{i+l+m}}{\sqrt{\xi_l\xi_m}}
              \frac{\sqrt{2}C_{ilms}^{\mu\nu\pm}(\ub{k_1},\ub{k_2})}
                          {\sqrt{1+\delta_{lm}\delta_{\mu\nu}}}\right|^2 .
  \nonumber
\eea
Using the coupled equations, we can express all the wave functions
$C_{ijs}^{\mu\pm}$ and $C_{ijks}^{\mu\nu\pm}$ and the amplitude
$z_1$ through the bare-electron amplitude $z_0$.
The normalization condition then determines $z_0$.  For the two-photon
truncation, where the wave functions are computed numerically, the
integrals for the normalization must also be done numerically, using
quadrature schemes discussed in Appendix~\ref{sec:quad}.

The anomalous moment $a_e$ can be computed from the spin-flip matrix element
of the electromagnetic current $J^+$~\cite{BrodskyDrell}
\be
-\left(\frac{Q_x-iQ_y}{2M}\right) F_2(Q^2)=
        \pm\frac12\langle \psi^\pm(\ub{P}+\ub{Q})| 
               \frac{J^+(0)}{P^+} |\psi^\mp(\ub{P})\rangle_{\rm phys} ,
\ee
where $Q$ is the momentum of the absorbed photon, $F_2$ is the
Pauli form factor, and we work in a frame where $Q^+$ is zero.
At zero momentum transfer, we have $a_e=F_2(0)$ and
\bea  \label{eq:TwoPhotonae}
a_e&=&m_e\sum_{s\mu}\int d\ub{k}\epsilon^\mu \sum_{j=0,2}\xi_j
  \left(\sum_{i'=0}^1\sum_{k'=j/2}^{j/2+1}
    \frac{(-1)^{i'+k'}}{\sqrt{\xi_{k'}}}C_{i'k's}^{\mu+}(\ub{k})\right)^* \\
  && \times y\left(\frac{\partial}{\partial k_x}+i\frac{\partial}{\partial k_y}\right)
  \left(\sum_{i=0}^1\sum_{k=j/2}^{j/2+1}
    \frac{(-1)^{i+k}}{\sqrt{\xi_k}}C_{iks}^{\mu-}(\ub{k})\right) \nonumber \\
  && +m_e \sum_{s\mu\nu}\int d\ub{k_1} d\ub{k_2} \sum_{j,k=0,2} \xi_j\xi_k  \nonumber \\
  && \rule{1in}{0in} \times
            \left( \sum_{i'=0}^1 
                \sum_{l'=j/2}^{j/2+1} \sum_{m'=k/2}^{k/2+1} 
                  \frac{(-1)^{i'+l'+m'}}{\sqrt{\xi_{l'}\xi_{m'}}}
              \frac{\sqrt{2}C_{i'l'm's}^{\mu\nu+}(\ub{k_1},\ub{k_2})}
                          {\sqrt{1+\delta_{l'm'}\delta_{\mu\nu}}}\right)^* \nonumber \\
   && \times \sum_a \left[y_a \left(\frac{\partial}{\partial k_{ax}}
                                   +i\frac{\partial}{\partial k_{ay}}\right)\right]
                 \left(\sum_{i=0}^1 
                \sum_{l=j/2}^{j/2+1} \sum_{m=k/2}^{k/2+1} 
                  \frac{(-1)^{i+l+m}}{\sqrt{\xi_l\xi_m}}
              \frac{\sqrt{2}C_{ilms}^{\mu\nu-}(\ub{k_1},\ub{k_2})}
                          {\sqrt{1+\delta_{lm}\delta_{\mu\nu}}}\right) .
                          \nonumber
\eea
In general, these integrals must also be computed numerically.

The terms that depend on the three-body wave functions $C_{ilms}^{\mu\nu\pm}$
are higher order in $\alpha$ than the leading two-body terms.  
This is because (\ref{eq:thirdcoupledequation}) determines $C_{ilms}^{\mu\nu\pm}$ as
being of order $\sqrt{\alpha}$ or $e$ times the two-body wave functions,
the vertex functions being proportional to the coupling, $e$.  Given the numerical 
errors in the leading terms, these three-body contributions are not significant
and are not evaluated.  The important three-body contributions come from the
couplings of the three-body wave functions that will enter the calculation 
of the two-body wave functions.

\section{Solution of the equations} \label{sec:twophoton}

\subsection{Integral equations for two-body wave functions} \label{sec:TwoPhotonIntegralEqns}

The first and third equations of the coupled system,
(\ref{eq:firstcoupledequation}) and (\ref{eq:thirdcoupledequation}),
can be solved for the bare-electron
amplitudes and one-electron/two-photon wave functions,
respectively, in terms of the one-electron/one-photon
wave functions.  From (\ref{eq:firstcoupledequation}), we have
\bea \label{eq:onebodyamplitude}
z_i & = & \frac{1}{M^2-m_i^2}\int d\ub{q} \sum_{j,l,\mu}\sqrt{\xi_l}(-1)^{j+l}\epsilon^\mu
  \left[P^+V_{ji\pm}^{\mu*}(\ub{P}-\ub{q},\ub{P})C^{\mu\pm}_{jl\pm}(\ub{q}) \right. \\
   && \left. \rule{2in}{0mm}
     +P^+U_{ji\pm}^{\mu*}(\ub{P}-\ub{q},\ub{P}) C^{\mu\pm}_{jl\mp}(\ub{q})\right],
                                        \nonumber
\eea
and from (\ref{eq:thirdcoupledequation}) we have
\bea \label{eq:threebodywavefunction}
C_{ijls}^{\mu\nu\pm}(\ub{q}_1,\ub{q}_2) & = & 
  \frac{1}{M^2 - \frac{m_i^2 + (\vec q_{1\perp}+\vec q_{2\perp})^2}{(1-y_1-y_2)} 
      - \frac{\mu_j^2 + q_{1\perp}^2}{y_1}  - \frac{\mu_l^2 + q_{2\perp}^2}{y_2} }
         \frac{\sqrt{1+\delta_{jl}\delta^{\mu\nu}}}{2} \\
    &&\times \sum_a (-1)^a 
      \left\{
        \sqrt{\xi_j}\left[V_{ias}^\mu(\ub{P}-\ub{q}_1-\ub{q}_2,\ub{P}-\ub{q}_2)
                              C_{als}^{\nu\pm}(\ub{q}_2)  \right.  \right.  \nonumber \\
     && \rule{1.25in}{0mm} \left.   +U_{ia,-s}^\mu(\ub{P}-\ub{q}_1-\ub{q}_2,\ub{P}-\ub{q}_2)
                              C_{al,-s}^{\nu\pm}(\ub{q}_2)\right]  \nonumber \\
      &&  \rule{1in}{0mm}
               +\sqrt{\xi_l}\left[V_{ias}^\nu(\ub{P}-\ub{q}_1-\ub{q}_2,\ub{P}-\ub{q}_1)
                              C_{ajs}^{\mu\pm}(\ub{q}_1) \right. \nonumber \\
       &&  \rule{1.25in}{0mm}
              \left. \left.  +U_{ia,-s}^\nu(\ub{P}-\ub{q}_1-\ub{q}_2,\ub{P}-\ub{q}_1)
                              C_{aj,-s}^{\mu\pm}(\ub{q}_1)\right] \right\} . \nonumber
\eea
Substitution of these solutions into
the second integral equation (\ref{eq:secondcoupledequation})
yields a reduced integral
eigenvalue problem in the one-electron/one-photon sector.

To isolate the dependence on the azimuthal angles, we use 
$q_i^1\pm iq_i^2=q_{i\perp} e^{\pm i\phi_i}$ and $q_i^+=y_iP^+$,
and write the vertex functions (\ref{eq:HelicityVertices}) as
\bea 
V_{ia\pm}^+(\ub{P}-\ub{q}_1-\ub{q}_2,\ub{P}-\ub{q}_2)&=&
   \frac{1}{(P^+)^{1/2}}\frac{e}{\sqrt{8\pi^3y_2}} , \\
V_{ia\pm}^-(\ub{P}-\ub{q}_1-\ub{q}_2,\ub{P}-\ub{q}_2)&=&
   \frac{1}{(P^+)^{5/2}}\frac{e}{\sqrt{8\pi^3y_2}}
             \frac{(q_{1\perp} e^{\mp i(\phi_1-\phi_2)}+q_{2\perp})q_{2\perp}+m_i m_a}
                   {(1-y_2)(1-y_1-y_2)} , \nonumber\\
V_{ia\pm}^{(\pm)}(\ub{P}-\ub{q}_1-\ub{q}_2,\ub{P}-\ub{q}_2)&=&
   -\frac{e^{\pm i\phi_2}}{(P^+)^{3/2}}\frac{e}{\sqrt{8\pi^3y_2}}
             \frac{q_{2\perp}}{1-y_2} ,\nonumber \\
V_{ia\mp}^{(\pm)}(\ub{P}-\ub{q}_1-\ub{q}_2,\ub{P}-\ub{q}_2)&=&
   -\frac{e^{\pm i\phi_2}}{(P^+)^{3/2}}\frac{e}{\sqrt{8\pi^3y_2}}
             \frac{q_{1\perp}e^{\pm i(\phi_1-\phi_2)}+q_{2\perp}}{1-y_1-y_2} ,  \nonumber\\
U_{ia\pm}^+(\ub{P}-\ub{q}_1-\ub{q}_2,\ub{P}-\ub{q}_2)&=&0 , \nonumber \\
U_{ia\pm}^-(\ub{P}-\ub{q}_1-\ub{q}_2,\ub{P}-\ub{q}_2)&=&
   \pm\frac{e^{\pm i\phi_2}}{(P^+)^{5/2}}\frac{e}{\sqrt{8\pi^3y_2}}
             \frac{m_a(q_{1\perp}e^{\pm i(\phi_1-\phi_2)}+q_{2\perp})-m_iq_{2\perp}}
                                {(1-y_2)(1-y_1-y_2)}  ,  \nonumber\\
U_{ia\pm}^{(\pm)}(\ub{P}-\ub{q}_1-\ub{q}_2,\ub{P}-\ub{q}_2)&=&0 , \nonumber \\
U_{ia\mp}^{(\pm)}(\ub{P}-\ub{q}_1-\ub{q}_2,\ub{P}-\ub{q}_2)&=&
   \mp\frac{1}{(P^+)^{3/2}}\frac{e}{\sqrt{8\pi^3y_2}}
             \frac{m_i (1-y_2)-m_a (1-y_1-y_2)}{(1-y_2)(1-y_1-y_2)} . \nonumber
\eea
The angular dependence of the wave functions is determined by the
sum of $J_z$ contributions for each Fock state.  For example, in the case
of $C_{ij-}^{(+)+}$, the photon is created by 
$a_{j(+)}^\dagger=\frac{1}{\sqrt{2}}(a_{j1}^\dagger-ia_{j2}^\dagger)$,
which contributes $J_z=-1$ to the state, and the constituent electron
contributes $J_z=-\frac12$; therefore, to have a total $J_z$ of $+\frac12$, the 
wave function must contribute $J_z=2$, which corresponds to a factor of
$e^{2i\phi}$.  For the full set of $J_z=+\frac12$ wave functions, we find
\bea
C_{ij+}^{++}(\ub{q})&=&\sqrt{P^+}C_{ij+}^{++}(y,q_\perp),\;\;
C_{ij+}^{-+}(\ub{q})=\frac{1}{P^{+3/2}}C_{ij+}^{-+}(y,q_\perp),\\
C_{ij+}^{(\pm)+}(\ub{q})&=&\frac{e^{\pm i\phi}}{\sqrt{P^+}}C_{ij+}^{(\pm)+}(y,q_\perp), 
\nonumber \\
C_{ij-}^{++}(\ub{q})&=&\sqrt{P^+}e^{i\phi}C_{ij-}^{++}(y,q_\perp),\;\;
C_{ij-}^{-+}(\ub{q})=\frac{e^{i\phi}}{P^{+3/2}}C_{ij-}^{-+}(y,q_\perp),\\
C_{ij-}^{(+)+}(\ub{q})&=&\frac{e^{2 i\phi}}{\sqrt{P^+}}C_{ij-}^{(+)+}(y,q_\perp), \;\;
C_{ij-}^{(-)+}(\ub{q})=\frac{1}{\sqrt{P^+}}C_{ij-}^{(-)+}(y,q_\perp). \nonumber
\eea
The wave functions have different dependence on longitudinal momenta,
resulting in different powers of $P^+$, which have been explicitly
factored out; they cancel against other $P^+$ factors in the final
integral equations.

The energy denominator of the three-body wave function can be written as
\bea
\lefteqn{M^2 - \frac{m_i^2 + (\vec q_{1\perp}+\vec q_{2\perp})^2}{(1-y_1-y_2)} 
      - \frac{\mu_j^2 + q_{1\perp}^2}{y_1}  - \frac{\mu_l^2 + q_{2\perp}^2}{y_2}}&& \\
    &&  =M^2 - \frac{m_i^2 + q_{1\perp}^2+q_{2\perp}^2 
              +2 q_{1\perp}q_{2\perp}\cos(\phi_1-\phi_2)}{(1-y_1-y_2)} 
      - \frac{\mu_j^2 + q_{1\perp}^2}{y_1}  - \frac{\mu_l^2 + q_{2\perp}^2}{y_2} . \nonumber
\eea
The light-cone volume element $d\ub{q}'$ becomes $\frac12 P^+ dy' d\phi' dq_\perp^{\prime 2}$.
All the angular dependence can then be gathered into integrals of the form
\be \label{eq:AngularIntegrals}
{\cal I}_n=\int_0^{2\pi} \frac{d\phi'}{2\pi}
 \frac{e^{in(\phi-\phi')}}{D_{ajb}(q_{1\perp},q_{2\perp})
                         +F(q_{1\perp},q_{2\perp})\cos(\phi-\phi')} ,
\ee
with $|n|=0,1,2,3$ and $D_{ajb}$ and $F$ defined as
\bea  \label{eq:FD}
D_{ajb}(q_\perp,q'_\perp)&=&\frac{m_a^2+q_\perp^2+q_\perp^{\prime 2}}{1-y-y'}
    +\frac{\mu_j^2+q_\perp^2}{y}+\frac{\mu_b^2+q_\perp^{\prime 2}}{y'}-M^2, \\
F(q_\perp,q'_\perp)&=&\frac{2q_\perp q'_\perp}{1-y-y'} . \nonumber
\eea

\begin{figure}[ht]
\vspace{0.1in}
\centerline{\includegraphics[width=12cm]{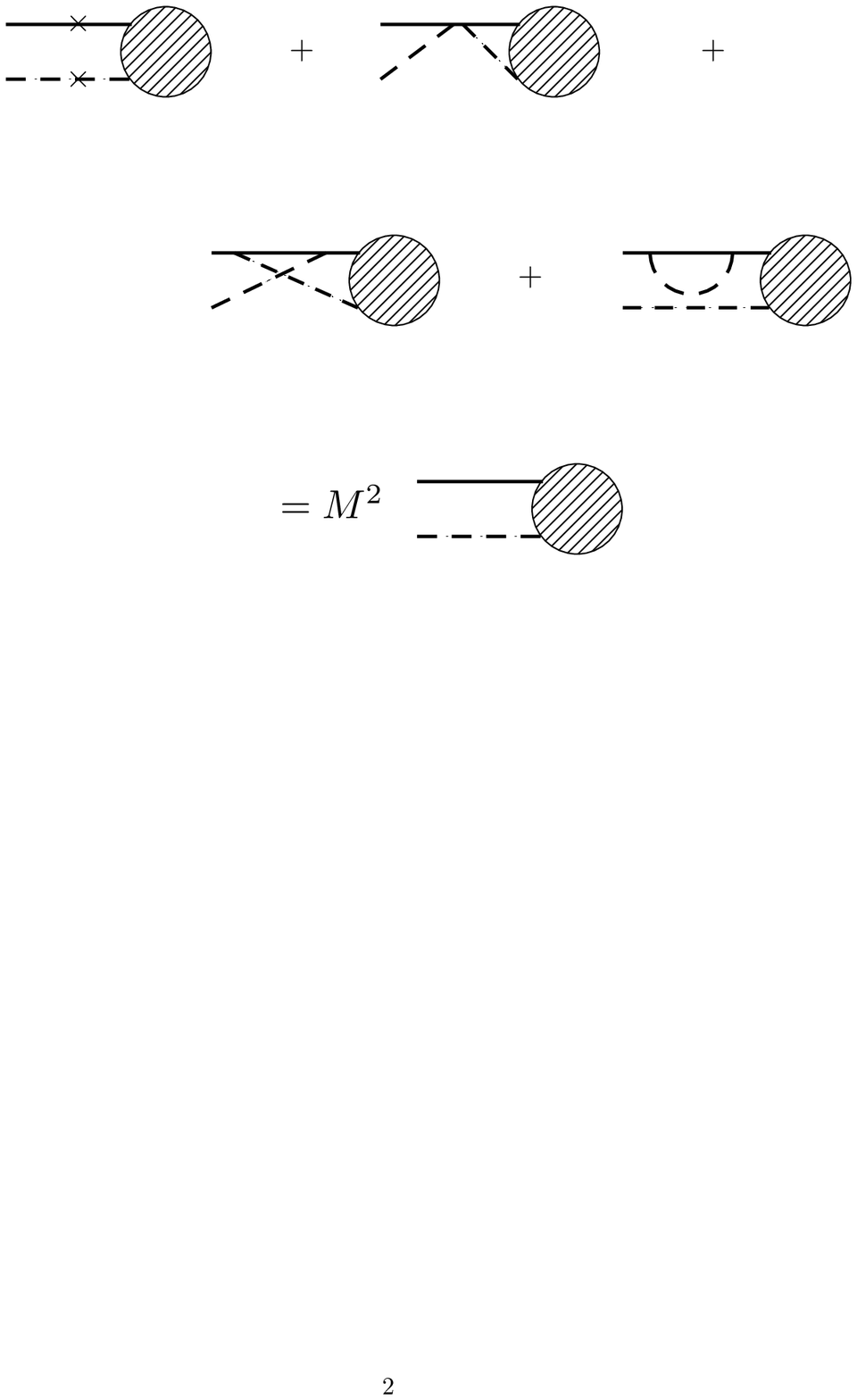}}
\caption[Diagrammatic representation of the coupled equations.]%
{\label{fig:viscacha2} Diagrammatic representation of
Eq.~(\ref{eq:ReducedEqn}) of the text.  The conventions for the
diagrams are the same as in Fig.~\protect\ref{fig:viscacha1}.}
\end{figure}

The integral equations for the two-body wave functions then take the form
\bea \label{eq:ReducedEqn}
\lefteqn{\left[M^2
  -\frac{m_i^2+q_\perp^2}{1-y}-\frac{\mu_j^2+q_\perp^2}{y}\right]
C_{ijs}^{\mu\pm}(y,q_\perp)=
\frac{\alpha}{2\pi}\sum_{i'}\frac{I_{iji'}(y,q_\perp)}{1-y}C_{i'js}^{\mu\pm}(y,q_\perp)}&& \\
 && \rule{1in}{0mm}
  +\frac{\alpha}{2\pi}\sum_{i'j's'\nu}\epsilon^\nu\int_0^1dy'dq_\perp^{\prime 2}
   J_{ijs,i'j's'}^{(0)\mu\nu}(y,q_\perp;y',q'_\perp)C_{i'j's'}^{\nu\pm}(y',q'_\perp) \nonumber \\
   &&  \rule{1in}{0mm}
   +\frac{\alpha}{2\pi}\sum_{i'j's'\nu}\epsilon^\nu\int_0^{1-y}dy'dq_\perp^{\prime 2}
   J_{ijs,i'j's'}^{(2)\mu\nu}(y,q_\perp;y',q'_\perp)C_{i'j's'}^{\nu\pm}(y',q'_\perp). \nonumber
\eea
There is a total of 48 coupled equations, with $i=0,1$; $j=0,1,2$; $s=\pm\frac12$;
and $\mu=\pm,(\pm)$.  
A diagrammatic representation is given in Fig.~\ref{fig:viscacha2}.%
%%%%%%%%%%%%%%%%%%%%%%%%%%%%%%%%%%%%%
%%%%%%%%%%%%%%%%%%%%%%%%%%%%%%%%%%%%%%%%%%%%%%%%%%%%%%%%%%%%%%

The first term on the right-hand side of (\ref{eq:ReducedEqn})
is the self-energy contribution~\cite{SecDep}:
\be  \label{eq:selfenergy}
I_{ili'}(y,q_\perp)
=\sum_{a,b}(-1)^{i'+a+b}\xi_b\int_0^1\frac{dx}{x}\frac{d^2k_\perp}{\pi}
\frac{m_i m_{i'} -2 \frac{m_i+m_{i'}}{1-x}m_a+\frac{m_a^2+k_\perp^2}{(1-x)^2}}
{\Lambda_l-\frac{m_a^2+k_\perp^2}{1-x}-\frac{\mu_b^2+k_\perp^2}{x}} ,
\ee
with
\be \label{eq:Lambda}
\Lambda_l\equiv \mu_l^2+(1-y)M^2-\frac{\mu_l^2+q_\perp^2}{y}.
\ee
The kernels $J^{(0)}$ and $J^{(2)}$ in the second and third terms
correspond to interactions with zero or two photons in intermediate states.  
The zero-photon kernel factorizes as
\be \label{eq:J0factorized}
J_{ijs,i'j's'}^{(0)\mu\nu}(y,q_\perp;y',q'_\perp)
  =\sum_a V_{ijas}^{(0)\mu}(y,q_\perp)\frac{(-1)^a}{M^2-m_a^2}V_{i'j'as'}^{(0)\nu*}(y',q'_\perp),
\ee
with
\be
V_{ija+}^{(0)+}=\sqrt{\xi_j}\frac{1}{\sqrt{y}},\;\;
V_{ija+}^{(0)-}=\sqrt{\xi_j}\frac{m_i m_a}{(1-y)\sqrt{y}}, \;\;
V_{ija+}^{(0)(+)}=0,\;\;
V_{ija+}^{(0)(-)}=\sqrt{\xi_j}\frac{q_\perp}{(1-y)\sqrt{y}},
\ee
and
\be
V_{ija-}^{(0)+}=0,\;\;
V_{ija-}^{(0)-}=\sqrt{\xi_j}\frac{m_a q_\perp}{(1-y)\sqrt{y}}, \;\;
V_{ija-}^{(0)(+)}=0,\;\;
V_{ija-}^{(0)(-)}=\sqrt{\xi_j}\frac{m_a(1-y)-m_i}{(1-y)\sqrt{y}} .
\ee
The two-photon kernels are considerably more involved, large in
number, and not particularly illuminating; details will not be
given here but can be found in \cite{thesis}.  The associated
angular integrals ${\cal I}_n$ are given in detail in
Appendix~\ref{sec:angularintegrals}.

\subsection{Fermion flavor mixing}

The presence of the flavor-changing self-energies,
the $I_{ili'}$ with $i\neq i'$, leads naturally to
a fermion flavor mixing of the two-body wave functions~\cite{SecDep}.
The integral equations for these functions have the structure
\bea \label{eq:MixingEquations}
A_{0j}C_{0js}^{\mu\pm} - B_jC_{1js}^{\mu\pm} &=& -\frac{\alpha}{2\pi}J_{0js}^{\mu\pm} , \\
B_j C_{0js}^{\mu\pm} + A_{1j} C_{1js}^{\mu\pm} &=& -\frac{\alpha}{2\pi} J_{1js}^{\mu\pm}, \nonumber
\eea
where $A_{ij}$ and $B_j$ are defined by
\be  \label{eq:Aij}
A_{ij}=\frac{m_i^2+q_\perp^2}{1-y}+\frac{\mu_j^2+q_\perp^2}{y}
    +\frac{\alpha}{2\pi}\frac{I_{iji}}{1-y}-M^2
\ee
and
\be \label{eq:Bj}
B_j=\frac{\alpha}{2\pi}\frac{I_{1j0}}{1-y}=-\frac{\alpha}{2\pi}\frac{I_{0j1}}{1-y},
\ee
and $J_{ijs}^{\mu\pm}$ is given by
\bea  \label{eq:Jmuijs}
J_{ijs}^{\mu\pm}&=&\sum_{i'j's'\nu}\epsilon^\nu\int_0^1dy'dq_\perp^{\prime 2}  
   J_{ijs,i'j's'}^{(0)\mu\nu}(y,q_\perp;y',q'_\perp)
            C_{i'j's'}^{\nu\pm}(y',q'_\perp) \\
  &&+\sum_{i'j's'\nu}\epsilon^\nu\int_0^{1-y}dy'dq_\perp^{\prime 2}
     J_{ijs,i'j's'}^{(2)\mu\nu}(y,q_\perp;y',q'_\perp)
            C_{i'j's'}^{\nu\pm}(y',q'_\perp) . \nonumber
\eea
The wave functions that diagonalize the left-hand side of 
(\ref{eq:MixingEquations}), and mix the physical ($i=0$) and 
PV ($i=1$) fermion flavors, are
\be
\tilde f_{ijs}^{\mu\pm}=A_{ij}C_{ijs}^{\mu\pm}+(-1)^i B_j C_{1-i,js}^{\mu\pm} .
\ee
In terms of these functions, the eigenvalue problem (\ref{eq:MixingEquations})
can be written as
\be
J_{ijs}^{\mu\pm}[\tilde f]=-\frac{2\pi}{\alpha}\tilde f_{ijs}^{\mu\pm} .
\ee
Here $J_{ijs}^{\mu\pm}$, the contribution of the zero-photon and two-photon
kernels, is implicitly a functional of these new wave functions.
The factors of $\alpha$ that appear in $A_{ij}$ and $B_j$ are assigned
the physical value and not treated as eigenvalues.
The original wave functions are recovered as
\be  \label{eq:orgwavefns}
C_{ijs}^{\mu\pm}=\frac{A_{1-i,js}\tilde f_{ijs}^{\mu\pm}
     +(-1)^i B_j \tilde f_{1-i,js}^{\mu\pm}}{A_{0j}A_{1j}+B_j^2}.
\ee
Self-energy contributions appear
in the denominators of the wave functions. 

To express the eigenvalue problem explicitly in terms of the
$\tilde f_{ijs}^{\mu\pm}$, we first write the definition
(\ref{eq:Jmuijs}) of $J_{ijs}^{\mu\pm}$ in a simpler form
\be
J_{ijs}^{\mu\pm}=\int dy' dq_\perp^{\prime 2}\sum_{i'j's'\nu}(-1)^{i'+j'}\epsilon^\nu
           J_{ijs,i'j's'}^{\mu\nu}(y,q_\perp;y',q'_\perp)C_{i'j's'}^{\nu\pm}(y',q'_\perp),
\ee
where $J_{ijs,i'j's'}^{\mu\nu}=J_{ijs,i'j's'}^{(0)\mu\nu}+J_{ijs,i'j's'}^{(2)\mu\nu}$.
Substitution of (\ref{eq:orgwavefns}) then yields, in matrix form,
\bea
\left(\begin{array}{c} J_{0js}^{\mu\pm} \\ J_{1js}^{\mu\pm} \end{array} \right)&=&
\int dy' dq_\perp^{\prime 2}\sum_{j's'\nu}(-1)^{j'}\epsilon^\nu
 \left(\begin{array}{cc} J_{0js,0j's'}^{\mu\nu} & J_{0js,1j's'}^{\mu\nu} \\
                         J_{1js,0j's'}^{\mu\nu} & J_{1js,1j's'}^{\mu\nu}
                                             \end{array}\right) \\
&& \rule{1.5in}{0mm} \times
 \left(\begin{array}{cc} A_{1j'} & B_{j'} \\
                         B_{j'} & -A_{0j'} \end{array}\right)
  \left(\begin{array}{c} \tilde f_{0j's'}^{\nu\pm} \\ \tilde f_{1j's'}^{\nu\pm} 
                                                            \end{array}\right). \nonumber
\eea
The sum over $\nu$ can also be written in matrix form for the
helicity components $\nu=\pm,\,(\pm)$ by the introduction of
\be
\lambda=\left(\begin{array}{cccc} 0 & -1 & 0 & 0 \\
                                  -1 & 0 & 0 & 0 \\
                                  0 & 0 & 1 & 0 \\
                                  0 & 0 & 0 & 1 \end{array} \right),
\ee
so that 
\be
\sum_{\nu}\epsilon^\nu J^{\mu\nu}\tilde f^{\nu\pm}=
  \sum_{\alpha,\beta}J^{\mu\alpha}\lambda_{\alpha\beta}\tilde f^{\beta\pm}.
\ee
Finally, we define
\be \label{eq:eta}
\eta_{j',\alpha\beta}=(-1)^{j'}\lambda_{\alpha\beta}
    \left(\begin{array}{cc} A_{1j'} & B_{j'} \\
                            B_{j'} & -A_{0j'} \end{array} \right)
\ee
as a tensor product of simpler matrices.  The eigenvalue problem
then becomes
\be \label{eq:fEqn}
\int dy' dq_\perp^{\prime 2}
      \sum_{i'j's'\alpha\beta i''} J_{ijs,i'j's'}^{\mu\alpha}(y,q_\perp;y',q'_\perp)
      \eta_{j',\alpha\beta,i'i''}\tilde f_{i''j's'}^{\beta\pm} 
      =-\frac{2\pi}{\alpha}\tilde f_{ijs}^{\mu\pm}.
\ee

This yields $\alpha$ as a function of $m_0$ and the PV masses.  We then find $m_0$
such that, for chosen values of the PV masses,
$\alpha$ takes the standard physical value $e^2/4\pi$. 
The eigenproblem solution also yields the functions $\tilde f_{ijs}^{\mu\pm}$ which
determine the wave functions $C_{ijs}^{\mu\pm}$.  From these wave functions
we can compute physical quantities as expectation values with respect to
the projection (\ref{eq:projected}) of the eigenstate onto the physical subspace.

\subsection{Numerical solution} \label{sec:numerical}

The eigenvalue problem (\ref{eq:fEqn}) is solved numerically.
Here we discuss the method used.
Additional details about quadratures can be found in Appendix~\ref{sec:quad}
and convergence properties are discussed in Appendix~\ref{sec:convergence}.

The integral equations (\ref{eq:fEqn}) for the wave functions 
of the electron are converted to a matrix eigenvalue problem by a discrete
approximation to the integrals, as discussed in Appendix~\ref{sec:quad}.
These approximations involve
variable transformations and Gauss--Legendre quadrature; the
transformations are done to minimize the number of quadrature
points required, in order to keep the matrix problem from
becoming too large, and to reduce the infinite transverse momentum 
range to a finite interval.  The resolution of the numerical
approximation is measured by two parameters, $K$ and $N_\perp$,
that control the number of quadrature points in the longitudinal
and transverse directions.

The integrals for the normalization
and anomalous moment, (\ref{eq:TwoPhotonNorm}) and 
(\ref{eq:TwoPhotonae}) respectively,
are also done numerically, but are summed over different
quadrature points.  These points take into account the different
shape of the integrand that comes from the square of the wave
functions.  The values of the wave functions at these
other points are found by cubic-spline interpolation~\cite{BurdenFaires}
in the transverse direction.
Regions of integration near the line of poles associated with the
energy denominator require special treatment, if the poles
exist, through quadrature formulas that take the
poles into account explicitly.

The renormalization requires finding the value of the
bare mass that corresponds to the physical value of
the coupling.  This defines a nonlinear equation for the
bare mass, which is solved with use of the M\"uller 
algorithm~\cite{BurdenFaires}.  Finding the poles in the
two-body wave function also requires solution of
nonlinear equations, and again the M\"uller algorithm 
is used.

The calculation of the anomalous moment requires computation
of a transverse derivative of the wave functions.  Because
the quadrature points used for integration are not uniformly
spaced, they are not convenient for  estimating the
derivative directly.  Instead, the wave functions are first
approximated by cubic splines; the derivatives are then obtained
from the splines.

To solve the eigenvalue problem, we treat the two-photon contributions explicitly,
but still nonperturbatively,
as corrections to the one-photon truncation with self-energy,
solved in \cite{SecDep}.  We do this by considering the coupled system
\bea \label{eq:onebodyAmp}
(M^2-m_a^2)z_a/z_0&=&\sqrt{\frac{\alpha}{2}} \sum_{i'j's'\alpha\beta i''}
   \int dy' dq_\perp^{\prime 2} V_{i'j'as'}^{(0)\alpha*}
         \eta_{j',\alpha\beta,i'i''} \tilde f_{i''j's'}^{\beta\pm}/z_0 , \\
\label{eq:twobodyWF}
\tilde f_{ijs}^{\mu\pm}/z_0&=&-\sqrt{\frac{\alpha}{2\pi^2}}
         \sum_a(-1)^aV_{ijas}^{(0)\mu}z_a/z_0 \\
&&     -\frac{\alpha}{2\pi}\int dy' dq_\perp^{\prime 2}
          \sum_{i'j's'\alpha\beta i''} J_{ijs,i'j's'}^{(2)\mu\alpha}
          \eta_{j',\alpha\beta,i'i''} \tilde f_{i''j's'}^{\beta\pm}/z_0, \nonumber
\eea
which can be obtained from (\ref{eq:onebodyamplitude}) and (\ref{eq:fEqn}),
with use of the factorization (\ref{eq:J0factorized}) for $J^{(0)}$ and
the connection (\ref{eq:orgwavefns}) between the original and flavor-mixed
two-body wave functions.

The solution is found by iteration.  When the index $a$ in 
(\ref{eq:onebodyAmp}) is equal to zero, we obtain an equation for $m_0$,
\be
m_0=+\sqrt{M^2-\sqrt{\frac{\alpha}{2}} \sum_{i'j's'\alpha\beta i''}
   \int dy' dq_\perp^{\prime 2} V_{i'j'0s'}^{(0)\alpha*}
         \eta_{j',\alpha\beta,i'i''} \tilde f_{i''j's'}^{\beta\pm}/z_0} ,
\ee
and when $a$ is equal to 1, we obtain an equation for $z_1$,
\be
z_1/z_0=\frac{1}{M^2-m_1^2}\sqrt{\frac{\alpha}{2}} \sum_{i'j's'\alpha\beta i''}
   \int dy' dq_\perp^{\prime 2} V_{i'j'1s'}^{(0)\alpha*}
         \eta_{j',\alpha\beta,i'i''} \tilde f_{i''j's'}^{\beta\pm}/z_0 .
\ee
These provide the updates of $m_0$ and $z_1/z_0$, and (\ref{eq:twobodyWF})
is solved by Jacobi iteration~\cite{BurdenFaires} of the linear system
that comes from the discretization of the rearrangement
\bea
\tilde f_{ijs}^{\mu\pm}/z_0
&+&\frac{\alpha}{2\pi}\int dy' dq_\perp^{\prime 2}
          \sum_{i'j's'\alpha\beta i''} J_{ijs,i'j's'}^{(2)\mu\alpha}
          \eta_{j',\alpha\beta,i'i''} \tilde f_{i''j's'}^{\beta\pm}/z_0 \\
&&=-\sqrt{\frac{\alpha}{2\pi^2}}
         \sum_a(-1)^aV_{ijas}^{(0)\mu}z_a/z_0 . \nonumber
\eea
Only a few Jacobi iterations are performed per update of $m_0$ and $z_1/z_0$;
further inner iteration is unnecessary, due to the subsequent changes in
$m_0$ and $z_1/z_0$.  The outer iterations of the full system of equations
is terminated when the changes in $m_0$, $z_1/z_0$, and the two-body
wave function are all of order $10^{-6}$ or less.  The bare amplitude
$z_0$ is obtained at the end by normalization.  The coupling $\alpha$ is held
fixed at the physical value; hence, this iterative method yields
not only the two-body wave functions and one-body amplitudes but also
the bare mass, $m_0$. 

The eigenvalue problem must first be solved for $M=0$, with the 
coupling strength parameter $\xi_2$ adjusted to yield $m_0=0$.  This
determines the value of $\xi_2$ that restores the chiral limit
nonperturbatively.  The eigenvalue problem can then be solved for
$M=m_e$, the physical mass of the electron, and the anomalous moment
calculated.

The number of wave-function updates in the Jacobi iteration
is small enough that the matrix
representing the discretization of the integral equations can be
computed at each iteration without making the calculation time
too large.  Thus, the matrix need not be stored, which allows much
higher resolutions.  We find that reasonable results are not
obtained until the longitudinal resolution $K$ is at least 50 and
that there is still considerable sensitivity to the longitudinal
resolution.  There is much less sensitivity to the transverse
resolution, for which $N_\perp=20$ is found sufficient.

\section{Results} \label{sec:results}

\begin{figure}[ht]
\centerline{\includegraphics[width=15cm]{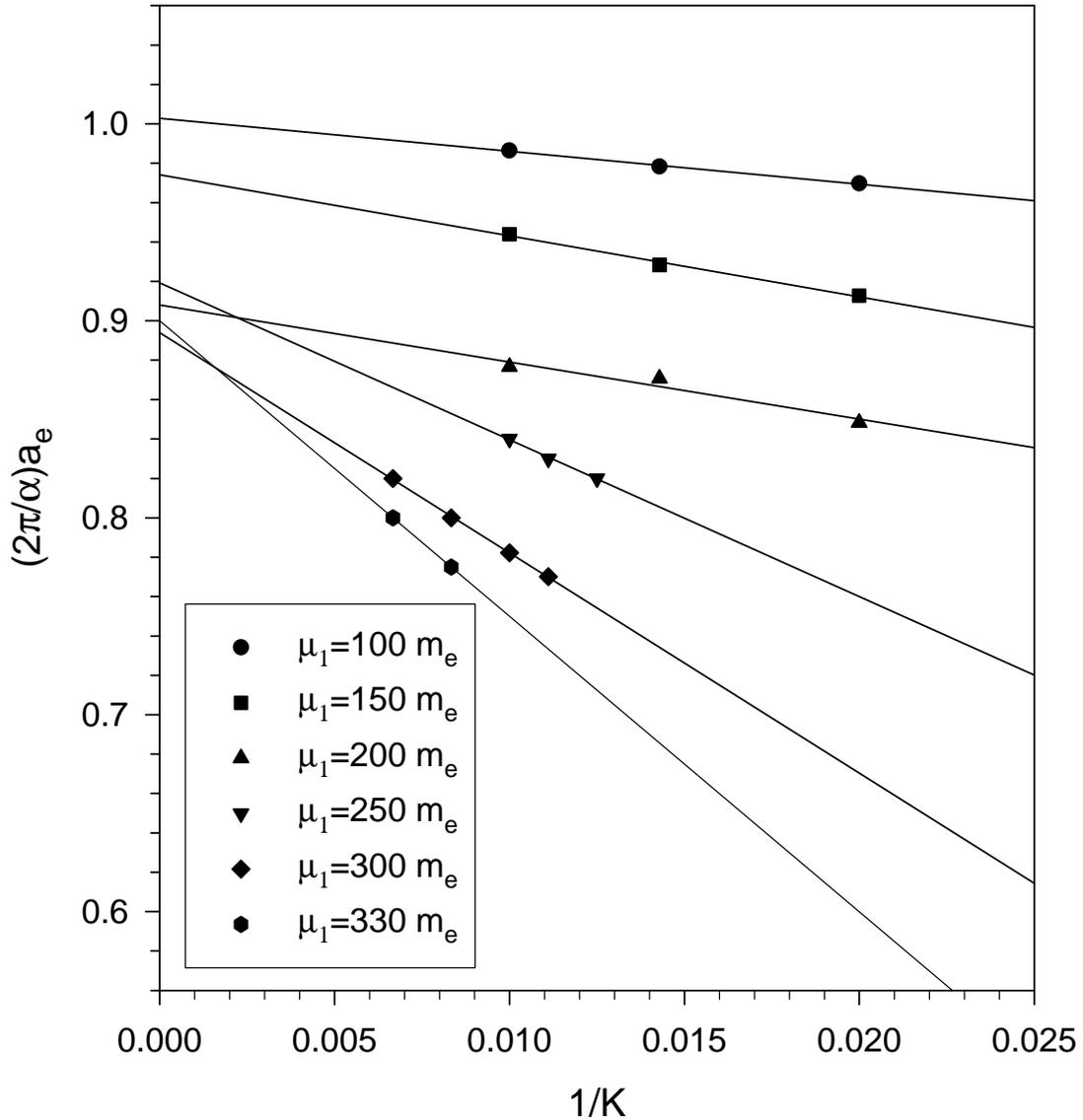}}
\caption{\label{fig:extrap} Dependence on longitudinal
resolution $K$ of the anomalous moment $a_e$ of the
electron in units of the Schwinger term ($\alpha/2\pi$)
for the two-photon truncation.
The PV masses are $m_1=2\cdot10^4m_e$, $\mu_1=100m_e$ to $330m_e$, 
and $\mu_2=\sqrt{2}\mu_1$.  The transverse resolution is $N_\perp=20$.
The lines are linear fits.  The errors in the individual points are
estimated to be 5\%.}
\end{figure}

Our results for particular values of longitudinal resolution $K$
are plotted in Fig.~\ref{fig:extrap}.  Several different values
are considered for the PV photon mass $\mu_1$, with $\mu_2$ fixed
as $\sqrt{2}\mu_1$ and the PV electron mass $m_1$ set to $2\cdot10^4m_e$.
The results are sensitive to $K$ even for these higher resolutions,
with greater sensitivity for the larger $\mu_1$ values.
In fact, beyond $\mu_1\simeq 300m_e$, convergence is difficult
to obtain for any value of $K$.

The choice of value for $\mu_2$ was studied in \cite{ChiralLimit}.
There is no particular sensitivity to the choice, provided $\mu_2$
is greater than $\mu_1$ and much less than $m_1$.  If $\mu_2$ is less
than $\mu_1$, the assignment of negative and positive metrics of
the two PV photons must be reversed.  If $\mu_2$ is too close to $m_1$,
observables can have a strong dependence on the PV masses.

We extrapolate the results for the anomalous moment
with linear fits in $1/K$.  The estimated
5\% error in the individual values, discussed in Appendix~\ref{sec:convergence},
does not justify a higher-order fit.  Given the nature of the fits,
we estimate an error of 10\% in the extrapolated values.

%%%%%%%%%%%%%%%%%%%%%%%%%%%%%%%%%%%%%
%%%%%%%%%%%%%%%%%%%%%%%%%%%%%%%%%%%%%%%%%%%%%%%%%%%%%%%%%%%%%%

\begin{figure}[ht]
\centerline{\includegraphics[width=15cm]{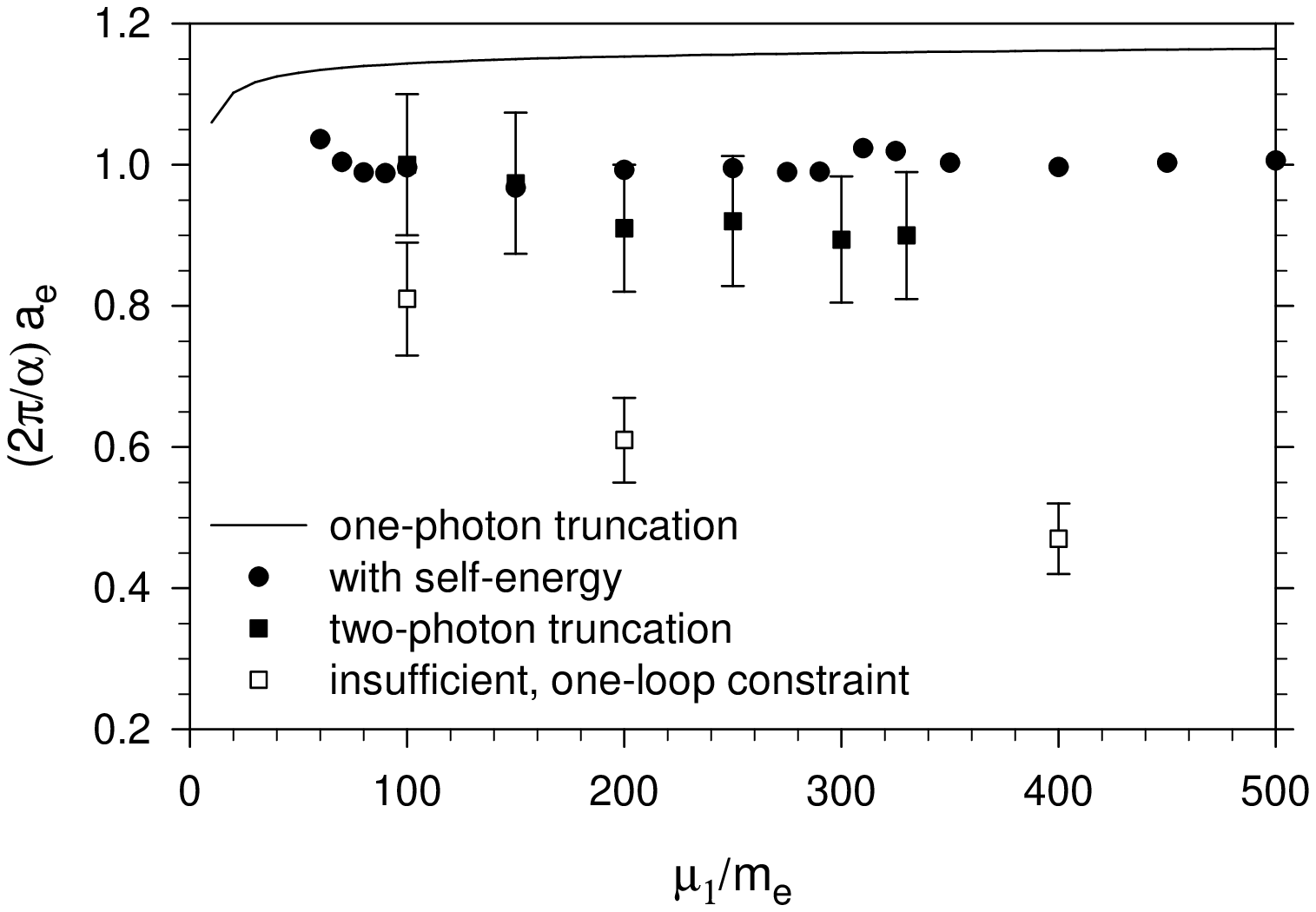}}
\caption{\label{fig:twophoton} The anomalous moment of the electron in
units of the Schwinger term ($\alpha/2\pi$) plotted versus
the PV photon mass, $\mu_1$, with the second PV photon mass, 
$\mu_2$, set to $\sqrt{2}\mu_1$
and the PV electron mass $m_1$ equal to $2\cdot10^4\,m_e$.
The solid squares are the result of the full two-photon truncation
with the correct, nonperturbative chiral constraint.  The open
squares come from use of a perturbative, one-loop constraint.
Results for the one-photon truncation~\protect\cite{ChiralLimit}
(solid line) and the one-photon truncation with the
two-photon self-energy contribution~\protect\cite{SecDep} (filled circles)
are included for comparison.
The resolutions used for the two-photon results are $K=50$ to 150,
combined with extrapolation to $K=\infty$, and $N_\perp=20$.}
\end{figure}

The results of the extrapolations are plotted in Fig.~\ref{fig:twophoton}.
Each value is close to the Schwinger result and independent of $\mu_1$, 
to within numerical error.  However, there is clearly a systematic tendency 
to be below the Schwinger result by approximately 10\% as the PV photon mass 
$\mu_1$ is increased.  We expect that this discrepancy is caused by the absence
of two potentially important contributions, the electron-positron loop
and the three-photon self-energy.  The loop contributes in perturbation
theory at the same order in $\alpha$ as the one-electron/two-photon
Fock states considered here and corresponds to the addition of
two-electron/one-positron Fock states to our truncation.  The
self-energy contribution is higher order in $\alpha$ but earlier
calculations~\cite{SecDep} have shown that the two-photon self-energy
is an important correction to the one-photon truncation.  This can be
seen in Fig.~\ref{fig:twophoton}, where we reproduce these results
for comparison.

Figure~\ref{fig:twophoton} also includes results obtained for the
two-photon truncation when only the one-loop chiral constraint
is satisfied.  Without the full
nonperturbative constraint, the results are very sensitive
to the PV photon mass $\mu_1$.  This behavior repeats the pattern observed
in \cite{ChiralLimit} for a one-photon truncation without the
corresponding one-loop constraint.  The resulting $\mu_1$
dependence is illustrated in Fig.~2 of \cite{ChiralLimit}.  Thus,
a successful calculation requires that the symmetry of the chiral
limit be maintained.

%%%%%%%%%%%%%%%%%%%%%%%%%%%%%%%%%%%%%
%%%%%%%%%%%%%%%%%%%%%%%%%%%%%%%%%%%%%%%%%%%%%%%%%%%%%%%%%%%%%%

%%%%%%%%%%%%%%%%%%%%%%
\section{Summary}
\label{sec:summary}
%%%%%%%%%%%%%%%%%%%%%%

The results of the calculation are shown in Fig.~\ref{fig:twophoton}
and compared with three other calculations: the one-photon 
truncation~\cite{ChiralLimit}, the one-photon case with the
two-photon self-energy contribution~\cite{SecDep}, and the
two-photon truncation with only the one-loop chiral constraint.
The two-photon results with the correct chiral constraint are
consistent with the Schwinger result, and therefore with experiment,
to within the estimated numerical error of 10\%.  The systematic
deviation below the Schwinger result is expected to
be due to the absence of
the two-electron/one-positron Fock sector and the three-photon
self-energy contributions. 
As is well known from perturbation theory, cancellations exist
between different types of contributions, such as between
photon loops and electron-positron loops, and, therefore, it
is not surprising for the present two-photon calculation, which
does not also include electron-positron loops, to have a
somewhat worse result than the one-photon calculation
with just the two-photon self-energy contribution.

Inclusion of an electron-positron
pair in the basis is also important for the understanding of
current covariance and of nonperturbative renormalization of
the charge and photon mass.
Future work along these lines will be to include these additional
contributions.  Of course, the calculation of the 
anomalous moment is not the important objective; instead, we are
interested in testing on QED a nonperturbative method
that could be applicable to QCD,
to see how the various truncations affect the result and to be
able to compare with perturbation theory, as a check.

Further tests of the method within QED could include application
to the calculation of a true bound state, positronium~\cite{Positronium}.
Just as for the anomalous moment calculation, the positronium
results will not
be competitive with high-order perturbation theory.
The numerical errors are large compared to the tiny perturbative
corrections in such a weakly coupled theory.  This in turn suggests 
another interesting test that could be done, a calculation of the
anomalous moment when $\alpha$ is much larger that its natural
value, yet small enough for perturbation theory to still provide
a check.

In a strongly coupled theory, such as QCD, the method may
be more quantitative.  For QCD, the PV-regulated formulation
by Paston {\em et al}.~\cite{Paston} could be a starting
point.  The analog of the dressed-electron problem does
not exist, of course, and the minimum truncation that
would include non-Abelian effects would be to include
at least two gluons.  The smallest calculation would
then be in the glueball sector.  In the meson sector, the
minimum truncation would be a quark-antiquark pair plus
two gluons, which as a four-body problem would require
discretization techniques beyond what are discussed here,
since the coupled integral equations for the wave functions
cannot be analytically reduced to a single Fock sector.
Instead, one would discretize the coupled integral equations
directly, in analogy with the original method of DLCQ~\cite{DLCQreview},
and diagonalize a very large but very sparse matrix.
As an intermediate step, one can select a less ambitious
yet very interesting challenge of modeling the meson
sector with effective interactions, particularly with
an interaction to break chiral symmetry~\cite{DalleyMcCartor}.

\acknowledgments
This work was supported in part by the Department of Energy
through Contract No.\ DE-FG02-98ER41087
and by the Minnesota Supercomputing Institute through
grants of computing time.

\appendix

\section{Discretizations and quadratures}  \label{sec:quad}

The integral equations involve integration over the longitudinal 
momentum fraction and the square of the transverse momentum.
The normalization, anomalous moment, and self-energy contributions
also require integrals of this form.  In each case there
can be a line of poles $q_{\rm pole}^2$ in the integrand,
from the denominators of wave functions in (\ref{eq:orgwavefns}), for 
a range of values of the longitudinal momentum fraction $y$.
The location of the line is determined by the energy denominator
that appears in each integrand.  For simple poles,
the transverse momentum integral is defined as 
the principal value.  For those values of longitudinal 
momentum $y$ for which the pole exists, the $q_\perp^2$
integration is subdivided into two parts, one from zero to
$2q_{\rm pole}^2$ and the other from there to infinity.  If
the pole does not exist, transverse integration is not 
subdivided.  When self-energy effects are included, the
location of the pole, if it still exists, must be found 
by solving a nonlinear equation numerically.  We do this
with the M\"uller algorithm~\cite{BurdenFaires}.

For the interval that contains a simple pole, the integral
is approximated by an open Newton--Cotes formula that
uses a few equally-spaced points placed symmetrically about the pole
at $q_i^2=(2i-1)q_{\rm pole}^2/N$ with $i=1,\ldots,N$ and $N$ even.
This particular Newton--Cotes formula uses a rectangular
approximation to the integrand, with the height equal to the
integrand value at the midpoint of an interval of width
$2q_{\rm pole}^2/N$.
An integral is then approximated by
\be
\int_0^{2q_{\rm pole}^2} dq_\perp^2 f(q_\perp^2)
\simeq \frac{2q_{\rm pole}^2}{N}\sum_{i=1}^N f(q_i^2).
\ee
The equally spaced points provide an approximation to the 
principal value.

This form avoids use of $q_\perp^2=0$ as a quadrature point.
Such a choice is important for evaluating terms with two-photon
kernels, where there is another pole associated with the
three-particle energy denominator.  By keeping $q_\perp^2$
nonzero, this pole can be handled analytically as
a principal value in the angular integration,
as discussed in Appendix~\ref{sec:angularintegrals}.

For the infinite intervals, $q_\perp^2$ is mapped to a new 
variable $v$ by the transformation~\cite{YukawaTwoBoson}
\be
q_\perp^2=a^2\frac{1-\left(b^2/a^2\right)^v}
                         {\left(b^2/a^2\right)^{v-1}-1},
\ee
with $v$ in the range 0 to 1.  (If the pole exists, this
transformation is shifted by $2q_{\rm pole}^2$.)  
The PV contributions make the integrals finite;
therefore, no transverse cutoff is needed.  Only the
positive Gauss--Legendre quadrature points of an even order $2N_\perp$
are used for $v$ between -1 and 1, so that $v=0$, and 
therefore $q'_\perp=0$ (or $2q_{\rm pole}^2$), is never a quadrature point.  
The points in the negative half of the range, which would be
used for representing $q_\perp^2\in[-\infty,0]$, are discarded.
One could map $q_\perp^2\in[0,\infty]$ to $[-1,1]$ and not
discard any part of the Gauss--Legendre range;
however, the quadrature would then place points focused on some
finite $q_\perp^2$ value, rather than on the natural integrand
peak at $q_\perp^2=0$.  The total number of quadrature points
in the transverse direction is $N_\perp+N$, with $N=0$ when there
is no pole and $N_\perp$ typically of order 20.

This transformation was used in \cite{YukawaTwoBoson}
and was selected to obtain an exact result
for the integral $\int [1/(a^2+q^2)-1/(b^2+q^2)]dq^2$.
In the present work, the scales $a^2$ and $b^2$ are chosen 
to be the smallest and largest scales in the problem, 
i.e.~$a^2=|q_{\rm pole}^2|$ and $b^2=m_1^2y+\mu_1^2(1-y)-M^2y(1-y)$.
Here $q_{\rm pole}^2$ is the location of the root of the nonlinear
equation for the pole.  If $q_{\rm pole}^2$ is negative, a pole
does not exist; however, $|q_{\rm pole}^2|$ is still a natural
scale for the integrand.

For the normalization and anomalous moment integrals, the
transverse quadrature scheme is based on a different
transformation
\be
q_\perp^2=a^2\frac{v}{1-v},
\ee
where, again, if the pole exists, the
transformation is shifted by $2q_{\rm pole}^2$.  Cubic-spline
interpolation~\cite{BurdenFaires}
is then used to compute the values of the
wave functions at the new quadrature points.  This
transformation is selected to yield an exact result
for the integral of $1/(a^2+q^2)^2$, which is the form of the
dominant contribution to the normalization and
anomalous moment.

The longitudinal integration is subdivided into three parts
when the line of poles is present.
Two parts are symmetrically placed about
the logarithmic singularity at $y_{\rm pole}$ that arises where the line
of poles reaches $q_\perp^2=0$.  When self-energy effects
are not included in the energy denominator, this occurs at
$y_{\rm pole}=1-m_0^2/M^2$;
when self-energy effects are present, the location must be
found by solving a nonlinear equation.  The third part of
the integration covers the remainder of the unit interval.  
Specifically, these intervals are $[0,y_{\rm pole}]$, $[y_{\rm pole},2y_{\rm pole}]$,
and $[2y_{\rm pole},1]$.
This structure is designed to maintain a left-right symmetry
around the logarithmic singularity, because in the 
normalization and anomalous moment integrals (which use
the same longitudinal quadrature points) the
singularity becomes a simple pole defined by a 
principal-value prescription.  The left-right symmetry
then assures the necessary cancellations from opposite
sides of the pole.  When no pole is present, the longitudinal
integration is not subdivided.

The intervals are each mapped linearly to $\tilde y\in[0,1]$ and
then altered by the transformations~\cite{YukawaTwoBoson}
\be \label{eq:ytilde}
\tilde y(t)=t^3(1+dt)/[1+d-(3+4d)t+(3+6d)t^2-4dt^3+2dt^4]
\ee
and
\be
t(u)=(u+1)/2.
\ee
The new variable $u$ ranges between -1 and 1, and standard
Gauss--Legendre quadrature is applied.  The transformation
from $\tilde y$ to $t$ is constructed to concentrate many
points near the end-points of each interval,
where integrands are rapidly varying.
The parameter $d$ is chosen such that $\tilde y\simeq 0.01 t^3$ for small $t$.
The transformation was found empirically~\cite{YukawaTwoBoson}, 
beginning with a transformation constructed to compute the
integral $\int_0^1[\ln(y+\epsilon_0)-\ln(y+\epsilon_1)] dy$ exactly,
with $\epsilon_0$ and $\epsilon_1$ small.  The symmetry with
respect to the replacements $t\rightarrow(1-t)$ and 
$\tilde y\rightarrow(1-\tilde y)$ is not necessary but is
the simplest choice for restricting the coefficients in the
denominator of (\ref{eq:ytilde}).

The need for a concentration of longitudinal quadrature points 
near 0 and 1 is particularly true for the integral $\bar J$, 
defined in (\ref{eq:J}).  Although this integral can be done
analytically for the case of the one-photon truncation discussed
in \cite{ChiralLimit}, the integral is only implicit in
the integral equations for the two-body wave functions discussed
in Sec.~\ref{sec:twophoton} and must therefore be well represented
by any discretization of the integral equations.  After the 
transverse integration is performed, the integrand is sharply
peaked near $y=0$ and $y=1$, at distances of order $m_0/m_1\sim10^{-10}$
from these end-points, and needs to be sampled on both sides
of the peaks.

The number of points in each of the three intervals is denoted
by $K$, which becomes the measure of the resolution analogous
to the harmonic resolution of DLCQ~\cite{PauliBrodsky}.
Thus the total number of quadrature points in the
longitudinal direction is $3K$, with $K$ typically
of order 50 or higher.

For those longitudinal integrals with an upper limit 
less than 1, the integrand is transformed as above
and given a value of zero for the points beyond the
original integration range.

For the normalization and anomalous moment integrals, 
the pole in the transverse integral (when it exists) is a
double pole, defined by the limit~\cite{OnePhotonQED}
\bea  \label{eq:doublepoledefinition}
   &&\int dy \; dq_\perp^2 \; 
   \frac{f(y,q_\perp^2)}{[ m^2 y + \mu_0^2 (1-y) -M^2 y (1-y) + q_\perp^2]^2}
   \nonumber \\
&&\equiv
 \lim_{\epsilon\rightarrow 0}
{\frac12 \epsilon} \int dy \int dq_\perp^2 f(y,k_\perp^2)
\Bigg[\frac{1}{[ m^2 y + \mu_0^2 (1-y) - M^2 y (1-y) + q_\perp^2 - \epsilon]}
   \nonumber \\
&&- \frac{1}{[ m^2 y + \mu_0^2 (1-y) - M^2 y (1-y) + q_\perp^2 +
\epsilon]}\Bigg].
\eea
The simple poles that remain are prescribed as principal values.  
Of course, the limit must be taken after the integral is performed.  

This limiting process is taken into account numerically by using a 
quadrature formula that is specific to this double-pole form.
On the interval $[0,2q_{\rm pole}^2]$,
the quadrature points are chosen to be the
same as those used for the integral equations, which are
$q_i^2=(2i-1)q_{\rm pole}^2/N$ with $i=1,\ldots,N$, as given above.  
The interval is divided into $N/2$ subintervals
$[\frac{4m}{N}q_{\rm pole}^2,\frac{4(m+1)}{N}q_{\rm pole}^2]$,
with $m=0,1,\ldots,(N-2)/2$, 
each containing two of the quadrature points.
The quadrature formula for such a subinterval is taken to be
\be
\int_{\frac{4m}{N}q_{\rm pole}^2}^{\frac{4(m+1)}{N}q_{\rm pole}^2}
dq_\perp^2 \frac{f(q_\perp^2)}{(q_\perp^2-q_{\rm pole}^2)^2}
\simeq  w_{2m+1} f(\frac{(4m+1)}{N}q_{\rm pole}^2) 
         +w_{2m+2} f(\frac{(4m+3)}{N}q_{\rm pole}^2),
\ee
where the integral on the left is {\em defined} by the limit
formula in (\ref{eq:doublepoledefinition}) when the pole
is in the subinterval.
The weights $w_i$ are chosen to make the formula exact
for $f=1$ and $f=q_\perp^2$ on each individual $q_\perp^2$ subinterval. 
For these numerator functions, the limit in (\ref{eq:doublepoledefinition})
can be taken explicitly.
The weights are then found to be
$w_{N/2}=w_{N/2+1}=-N/2q_{\rm pole}^2$, for the quadrature points 
on either side of the pole.  For all other points, the weights are given by
\bea
w_{2m+1}&=&-\frac{N}{2q_{\rm pole}^2}\left[\ln\left|\frac{4m+4-N}{4m-N}\right|
  +\frac{4(N-4m-3)}{(4m-N)(4m+4-N)}\right], \\
w_{2m+2}&=&\frac{N}{2q_{\rm pole}^2}\left[\ln\left|\frac{4m+4-N}{4m-N}\right|
  +\frac{4(N-4m-1)}{(4m-N)(4m+4-N)}\right].  \nonumber
\eea
The integral from 0 to $2q_{\rm pole}^2$ is obtained by summing
over the individual subintervals.

For the self-energy contribution (\ref{eq:selfenergy}), which is expressed in
terms of the integrals $\bar I_0$, $\bar I_1$, and $\bar J= M^2 \bar I_0$
given in (\ref{eq:In}) and (\ref{eq:J}),
the transverse integral is done analytically.  Only the longitudinal
integral is done numerically, by the scheme discussed above with 
resolution $K=30$.

\section{Angular integrals} \label{sec:angularintegrals}

Calculation of the two-photon kernels requires the integrals
\be \label{eq:AngInt}
{\cal I}_n=\int_0^{2\pi} \frac{d\phi}{2\pi}
 \frac{e^{-in\phi}}{D+F\cos\phi} ,
\ee
first defined in Eq.~(\ref{eq:AngularIntegrals}),
with $D$ and $F$ given in (\ref{eq:FD}).  
Here the original integration variable $\phi'$ has been
shifted by the independent angle $\phi$, and the prime then
dropped for simplicity of notation in this Appendix.
The factor $F$ is always positive, but $D$ can be negative.
If the bare fermion mass $m_0$ is less than
the physical mass $m_e$, we can have $|D|<F$; in this case, ${\cal I}_n$ is
defined by a principal value, as in the one-photon sector.  If either photon
has zero transverse momentum, $F$ will be zero, and any pole due to a zero
in $D$ will not involve the angular integration.  The numerical quadrature
is chosen to never use grid points where a photon transverse momentum
is zero, so that the principal-value prescription can always be invoked 
for the angular integral, where it is easily handled analytically.

The imaginary part of ${\cal I}_n$ is zero.  This follows from the
even parity of the denominator and the odd parity of $\sin n\phi$.
As a consequence, ${\cal I}_{-n}={\cal I}_n$, and we evaluate 
(\ref{eq:AngInt}) for only nonnegative $n$.

The real part is nonzero and most easily calculated from combinations
of the related integrals
\be
\bar{\cal I}_n=\int_0^{2\pi} \frac{d\phi}{2\pi} \frac{\cos^n\phi}{D+F\cos\phi}.
\ee
Of course, for $n=0$ and 1, the two integrals are identical.  For $n=2$ and
3 we have $\cos 2\phi=2\cos^2\phi-1$ and $\cos 3\phi=4\cos^3\phi-3\cos\phi$.
Therefore, the integral combinations are
\be
{\cal I}_0=\bar{\cal I}_0 , \;\;
{\cal I}_1=\bar{\cal I}_1 , \;\;
{\cal I}_2=2\bar{\cal I}_2-\bar{\cal I}_0, \;\;
{\cal I}_3=4\bar{\cal I}_3-3\bar{\cal I}_1 .
\ee
Larger values of $n$ do not appear in the two-photon kernels.

The integrals $\bar{\cal I}_n$ are connected by a simple recursion for $n>0$:
\be \label{eq:calIrecursion}
\bar{\cal I}_n=\int_0^{2\pi} \frac{d\phi}{2\pi}\frac{\cos^{n-1}\phi}{F}
      \frac{(D+F\cos\phi-D)}{D+F\cos\phi}
      =\frac{1}{F}\int_0^{2\pi} \frac{d\phi}{2\pi}\cos^{n-1}\phi
                                 -\frac{D}{F}\bar{\cal I}_{n-1} .
\ee
The first term is zero when $n$ is even.  For $n=1$, it is $1/F$, and for
$n=3$, this term is $1/2F$.  The only other integral that must be evaluated
directly is $\bar{\cal I}_0={\cal I}_0$.

\begin{figure}[ht]
\centerline{\includegraphics[width=9cm]{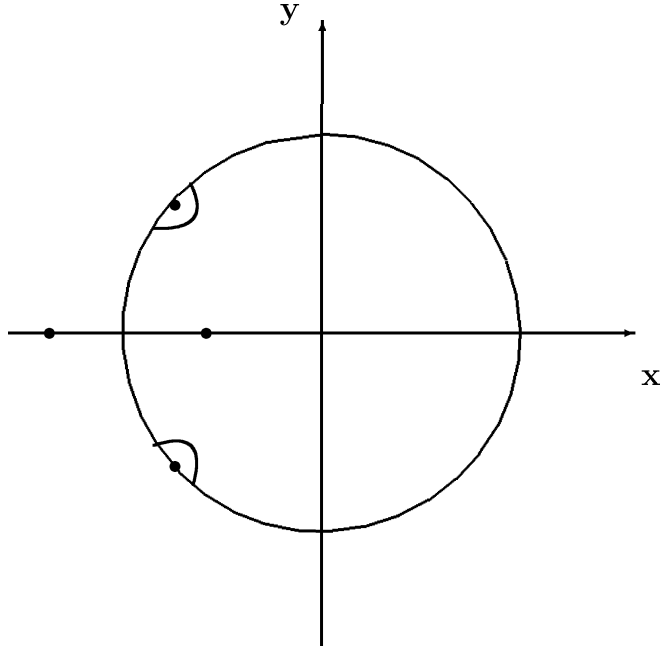}}
%\centerline{\input{contour.pic}}
\caption[Integration contour for evaluation
of ${\cal I}_0$.]%
{\label{fig:contour} Integration contour for evaluation
of ${\cal I}_0$.  The locations of the poles at $z_\pm$ depend
upon the magnitude and sign of $D/F$.  The semicircles are used
when $|D|/F<1$ and the poles are on the contour.
}
\end{figure}

The determination of ${\cal I}_0$, with or without the presence
of poles, is conveniently done by contour integration around the unit
circle in terms of a complex variable $z=e^{i\phi}$.  We then have
\be
{\cal I}_0=\frac{1}{i\pi F}\oint\frac{dz}{z^2+2\frac{D}{F}z+1}
   =\frac{1}{i\pi F}\oint \frac{dz}{(z-z_+)(z-z_-)} .
\ee
There are simple poles at
\be
z_\pm=-\frac{D}{F}\pm\sqrt{\frac{D^2}{F^2}-1}
       =-\frac{D}{F}\pm i\sqrt{1-\frac{D^2}{F^2}}
       =-e^{\mp i\cos^{-1}(D/F)}
\ee
When $D$ is greater than $F$, one pole, $z_+$, is inside the contour
and the other outside, as illustrated in Fig.~\ref{fig:contour}.
%%%%%%%%%%%%%%%%%%%%%%%%%%%%%%%%%%%%%
%%%%%%%%%%%%%%%%%%%%%%%%%%%%%%%%%%%%%%%%%%%%%%%%%%%%%%%%%%%%%%
Evaluation of $2\pi i$ times the residue yields
\be
{\cal I}_0=\frac{1}{\sqrt{D^2-F^2}} \;\; \mbox{for} \;\; D>F .
\ee
Similarly, when $D$ is less than $-F$, the pole at $z_-$ is inside
the contour, and we have
\be
{\cal I}_0=-\frac{1}{\sqrt{D^2-F^2}} \;\; \mbox{for} \;\; D<-F .
\ee
When $|D|$ is less than $F$, the poles move to the contour, and the 
integral is defined by the principal value.  This is evaluated
by distorting the contour to include semicircles of radius $\epsilon$
around each pole, as shown in Fig.~\ref{fig:contour}, and subtracting
the contributions from the semicircles after taking the 
$\epsilon\rightarrow0$ limit.  The choice of inward semicircles
makes the integral around the closed contour simply zero.
For the semicircle around $z_\pm$, we have $z=z_\pm+\epsilon e^{i\theta}$
and a contribution, as $\epsilon$ goes to zero, of
\be
\frac{2}{iF}\int 
   \frac{\epsilon i e^{i\theta} d\theta}
         {\epsilon e^{i\theta} (\epsilon e^{i\theta}\mp 2i\sqrt{1-D^2/F^2})}
      \longrightarrow
      \frac{\pm\int d\theta}{F\sqrt{1-D^2/F^2}}.
\ee
Thus the contributions from the two semicircles are of opposite sign
and cancel, so that the net result is also zero.  Therefore, we have
\be
{\cal I}_0=\left\{ \begin{array}{ll} \frac{1}{\sqrt{D^2-F^2}} , & D>F \\
                                        0 , & |D|<F \\
                                     -\frac{1}{\sqrt{D^2-F^2}} , & D<-F .
                                     \end{array}\right.
\ee
The case where $D$ equals $F$ represents an integrable singularity
for the transverse momentum integrations and can be ignored.  When $F$
is zero, we have simply
\be
{\cal I}_n=\int_0^{2\pi} \frac{d\phi }{2\pi D}e^{-in\phi}=\frac1D\delta_{n0}.
\ee

When $F/D$ is small, the expressions for the integrals 
${\cal I}_n$ are best evaluated from expansions in powers
of $F/D$, to avoid round-off errors due to cancellations between
large contributions.  The expansions used are
\bea
{\cal I}_0&\simeq&\frac{1}{128D}\left[128+64\left(\frac{F}{D}\right)^2
            + 48\left(\frac{F}{D}\right)^4    
               + 40\left(\frac{F}{D}\right)^6 + 35\left(\frac{F}{D}\right)^8\right] , \\ \nonumber
{\cal I}_1&\simeq&-\frac{1}{128D}\left(\frac{F}{D}\right)
     \left[64 + 48\left(\frac{F}{D}\right)^2 + 40\left(\frac{F}{D}\right)^4
              + 35\left(\frac{F}{D}\right)^6\right] , \\  \nonumber
{\cal I}_2&\simeq&\frac{1}{128D}\left(\frac{F}{D}\right)^2\left[32+32\left(\frac{F}{D}\right)^2
                  +30\left(\frac{F}{D}\right)^4\right] , \\ \nonumber
{\cal I}_3&\simeq&-\frac{1}{128D}\left(\frac{F}{D}\right)^3
       \left[16+20\left(\frac{F}{D}\right)^2\right]  .
\eea

\section{Numerical convergence}  \label{sec:convergence}

The primary constraint on numerical accuracy is the error
in the estimation of the integrals in the integral equations
for the wave functions and in the expressions for the 
normalization and the anomalous moment.  This accuracy
is determined by the choice of quadrature scheme,
discussed in Appendix~\ref{sec:quad}, and the resolution,
controlled by the longitudinal parameter $K$ and transverse
parameter $N_\perp$.  The other numerical parts of the
calculation are iterated to what is effectively
exact convergence, with remaining uncertainties much smaller
than the errors in the numerical quadratures.

%%%%%%%%%%%%%%%%%%%%%%%%%%%%%%%%%%%%%
\begin{table}[ht]
\caption[Dependence on longitudinal
resolution of the integrals $I_0$, $I_1$, and $J$.]%
{\label{tab:I0I1JvsK} Dependence on longitudinal
resolution $K$ of the integrals $\bar I_0$, $\bar I_1$, and $\bar J$, defined
in (\ref{eq:In}) and (\ref{eq:J}) of the text and computed
according to the quadrature scheme described in Appendix~\ref{sec:quad}.
The bare-electron mass is $m_0=0.98m_e$.
The PV masses are $m_1=2\cdot10^4m_e$, $\mu_1=200m_e$,
and $\mu_2=\sqrt{2}\mu_1$.  The transverse resolution
is $N_\perp=40$.}
\begin{tabular}{r|ccc}
\hline \hline
$K$  &  $\bar I_0(m_e^2)$ & $\bar I_1(m_e^2)/m_e$ & $\bar J(m_e^2)/m_e^2$ \\
\hline
5 & -7.113 & -13.530 & -11606. \\
10 & -6.2586 & -10.6182 & 932.3 \\
15 & -6.2641 & -10.7354 & -26.487 \\
20 & -6.2645 & -10.7327 & -6.7126 \\
25 & -6.2645 & -10.7328 & -6.3982 \\
30 & -6.2645 & -10.7328 & -6.4401 \\
\hline
exact & -6.2645 & -10.7328 & -6.2645 \\
\hline
\hline
\end{tabular}
\end{table}
%%%%%%%%%%%%%%%%%%%%%%%%%%%%%%%%%%%%%%%%%%%%%%%%%%%%%%%%%%%%%%
%
%%%%%%%%%%%%%%%%%%%%%%%%%%%%%%%%%%%%%
\begin{figure}[ht]
\vspace{0.1in}
\centerline{\includegraphics[width=15cm]{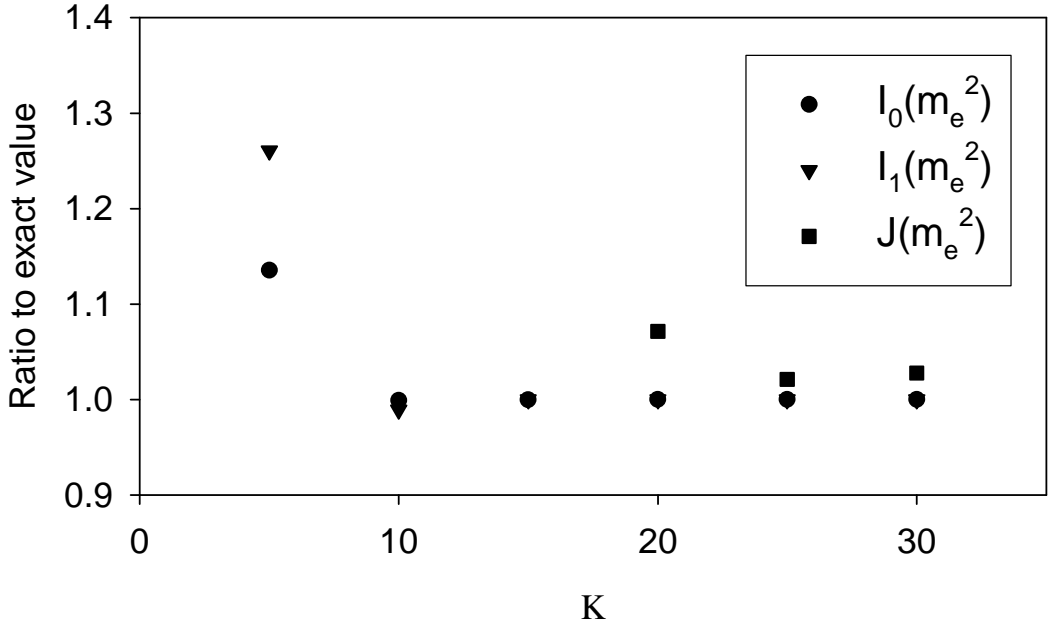}}
\caption[Dependence on longitudinal
resolution of the integrals $I_0$, $I_1$, and $J$.]%
{\label{fig:I0I1JvsK} Dependence on longitudinal
resolution $K$ of the integrals $I_0$, $I_1$, and $J$, defined
in (\ref{eq:In}) and (\ref{eq:J}) of the text and computed
according to the quadrature scheme described in Appendix~\ref{sec:quad}.
The values plotted are ratios to the exact values, listed in Table~\ref{tab:I0I1JvsK}.
The bare-electron mass is $m_0=0.98m_e$.
The PV masses are $m_1=2\cdot10^4m_e$, $\mu_1=200m_e$,
and $\mu_2=\sqrt{2}\mu_1$.  The transverse resolution
is $N_\perp=40$.}
\end{figure}
%%%%%%%%%%%%%%%%%%%%%%%%%%%%%%%%%%%%%%%%%%%%%%%%%%%%%%%%%%%%%%
%
%%%%%%%%%%%%%%%%%%%%%%%%%%%%%%%%%%%%%
\begin{table}[ht]
\caption[Same as Table~\ref{tab:I0I1JvsK}, but
for the dependence on transverse resolution.]%
{\label{tab:I0I1JvsNperp} Same as Table~\ref{tab:I0I1JvsK}, but
for the dependence on transverse resolution $N_\perp$.
The longitudinal resolution is $K=30$.}
\begin{tabular}{r|ccc}
\hline \hline
$N_\perp$  &  $\bar I_0(m_e^2)$ & $\bar I_1(m_e^2)/m_e$ & $\bar J(m_e^2)/m_e^2$ \\
\hline
10 & -6.2646 & -10.7324 & -2.9845 \\
15 & -6.2645 & -10.7327 & -7.5428 \\
20 & -6.2645 & -10.7327 & -6.9137 \\
25 & -6.2645 & -10.7328 & -6.6961 \\
30 & -6.2645 & -10.7328 & -6.5712 \\
35 & -6.2645 & -10.7328 & -6.4922 \\
40 & -6.2645 & -10.7328 & -6.4401 \\
\hline
exact & -6.2645 & -10.7328 & -6.2645 \\
\hline
\hline
\end{tabular}
\end{table}
%%%%%%%%%%%%%%%%%%%%%%%%%%%%%%%%%%%%%%%%%%%%%%%%%%%%%%%%%%%%%%
%
%%%%%%%%%%%%%%%%%%%%%%%%%%%%%%%%%%%%%
\begin{figure}[ht]
\vspace{0.1in}
\centerline{\includegraphics[width=15cm]{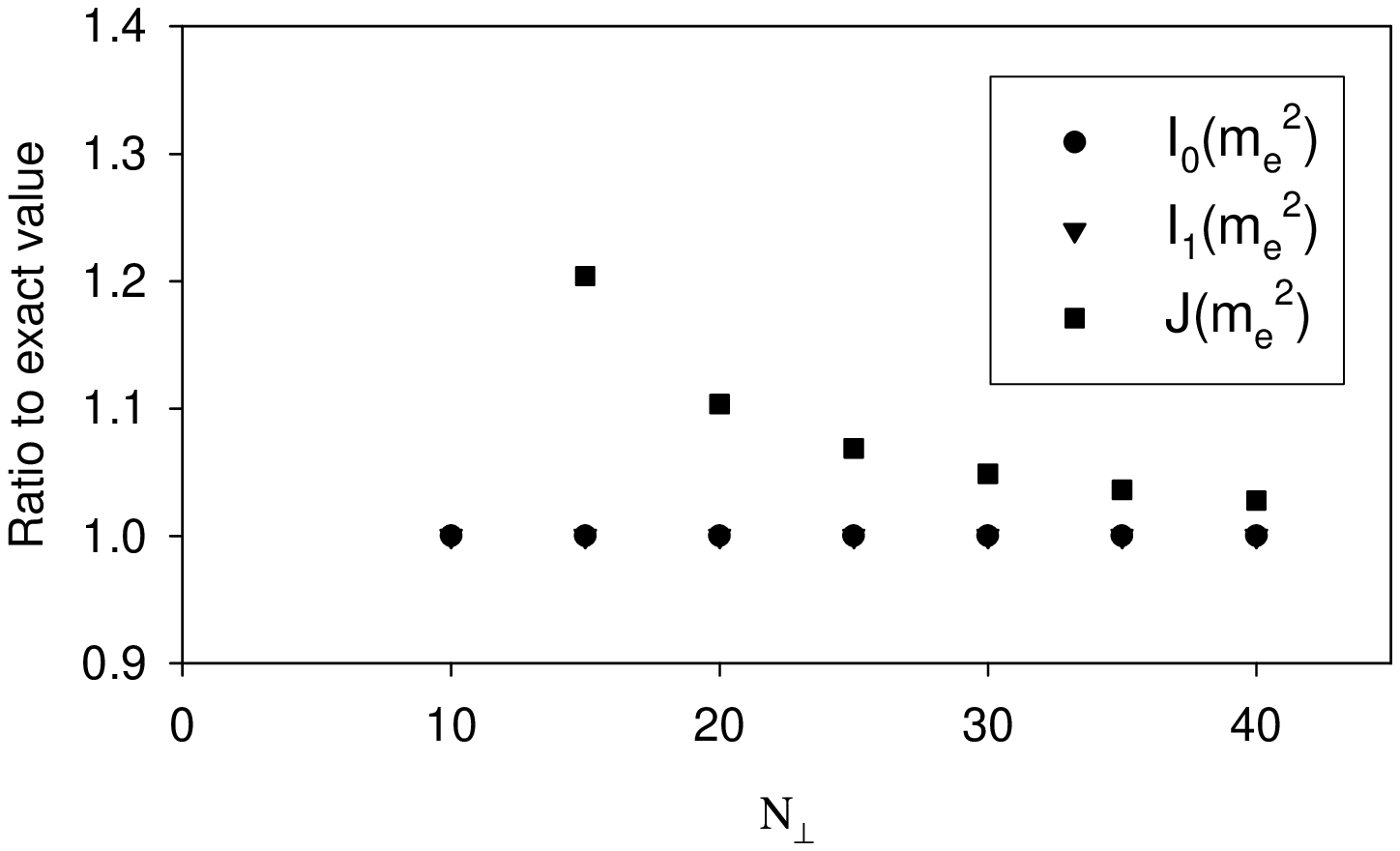}}
\caption[Same as Fig.~\ref{fig:I0I1JvsK}, but
for the dependence on transverse resolution.]%
{\label{fig:I0I1JvsNperp} 
Same as Fig.~\ref{fig:I0I1JvsK}, but
for the dependence on transverse resolution $N_\perp$.
The longitudinal resolution is $K=30$.
For these resolutions, the values for $\bar I_0$ and $\bar I_1$
are nearly exact, and the plotted points for the ratios to the
exact values are at the same places; only $J$ shows variation.}
\end{figure}
%%%%%%%%%%%%%%%%%%%%%%%%%%%%%%%%%%%%%%%%%%%%%%%%%%%%%%%%%%%%%%
%

For the one-photon truncation, discussed in \cite{OnePhotonQED} 
and \cite{ChiralLimit},
all the integrals can be done analytically.  This makes the
one-photon problem a convenient first test for numerical
convergence.  The key integrals are $\bar I_0$, $\bar I_1$,
and $\bar J$, defined as
\begin{eqnarray} \label{eq:In}
\bar{I}_n(M^2)&=&\int\frac{dy dk_\perp^2}{16\pi^2}
   \sum_{jl}\frac{(-1)^{j+l}\xi_l}{M^2-\frac{m_j^2+k_\perp^2}{1-y}
                                   -\frac{\mu_l^2+k_\perp^2}{y}}
   \frac{m_j^n}{y(1-y)^n}\,, \\
\bar{J}(M^2)&=&\int\frac{dy dk_\perp^2}{16\pi^2}  \label{eq:J}
   \sum_{jl}\frac{(-1)^{j+l}\xi_l}{M^2-\frac{m_j^2+k_\perp^2}{1-y}
                                   -\frac{\mu_l^2+k_\perp^2}{y}}
   \frac{m_j^2+k_\perp^2}{y(1-y)^2} .
\end{eqnarray}
Tables~\ref{tab:I0I1JvsK} and \ref{tab:I0I1JvsNperp}
and Figs.~\ref{fig:I0I1JvsK} and \ref{fig:I0I1JvsNperp}
summarize results for numerical calculation of these
integrals.  They show that $\bar I_0$ and $\bar I_1$ are well
approximated for a wide range of resolutions, but $\bar J$ is
particularly sensitive to the longitudinal resolution $K$
and requires that both $K$ and $N_\perp$ be on the order
of 20 or larger.  At these resolutions, $\bar J$ is approximated
with an accuracy of about 4\%, and this then becomes a
minimal estimate of the accuracy of any of the results.

%
%%%%%%%%%%%%%%%%%%%%%%%%%%%%%%%%%%%%%
\begin{table}[ht]
\caption[Dependence on longitudinal
resolution of the anomalous moment.]%
{\label{tab:aevsK} Dependence on longitudinal
resolution $K$ of the bare mass $m_0$ and anomalous moment $a_e$
of the electron in units of the physical mass $m_e$ and
the Schwinger term ($\alpha/2\pi$), respectively,
for both the one-photon truncation, when solved numerically,
and the case with the self-energy included.
The value of the bare mass is obtained as the value that
yields a physical value of the coupling constant $\alpha$;
for $K=15$, two solutions are found.
The PV masses are $m_1=2\cdot10^4m_e$, $\mu_1=200m_e$,
and $\mu_2=\sqrt{2}\mu_1$.  The transverse resolution
is $N_\perp=40$.}
\begin{tabular}{r|cc|cc}
\hline \hline
      & \multicolumn{2}{c|}{one-photon} & \multicolumn{2}{c|}{with self-energy} \\
\cline{2-5}
$K$ & $m_0/m_e$ & $2\pi a_e/\alpha$ & $m_0/m_e$ & $2\pi a_e/\alpha$   \\
\hline
5 & 4.4849 & 0.09921 & 4.4845 & 0.09444 \\
10 & 1.7445 & 0.22992 & 1.7106 & 0.22620 \\
15  & 1.07487 & 0.64482 & 1.07481 & 0.59092 \\
15 & 0.98906 & 1.12763 & 0.97537 & 1.07444 \\
20 & 0.98223 & 1.15430 & 0.99028 & 0.90337 \\
25 & 0.98240 & 1.15382 & 0.98295 & 0.99242 \\
30 & 0.98241 & 1.15567 & 0.98245 & 1.00612 \\
\hline \hline
\end{tabular}
\end{table}
%%%%%%%%%%%%%%%%%%%%%%%%%%%%%%%%%%%%%%%%%%%%%%%%%%%%%%%%%%%%%%
%
%%%%%%%%%%%%%%%%%%%%%%%%%%%%%%%%%%%%%
\begin{figure}[ht]
\centerline{\includegraphics[width=15cm]{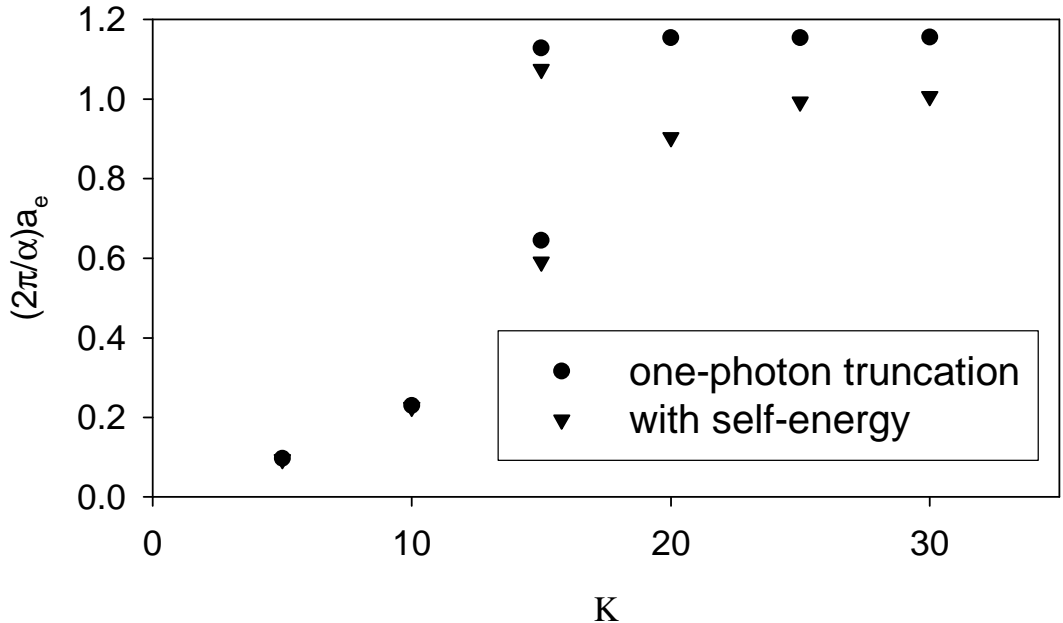}}
\caption[Dependence on longitudinal
resolution of the anomalous moment.]%
{\label{fig:aevsK} Dependence on longitudinal
resolution $K$ of the anomalous moment $a_e$ of the
electron in units of the Schwinger term ($\alpha/2\pi$)
for both the one-photon truncation, when solved numerically,
and the case with the self-energy included.
The PV masses are $m_1=2\cdot10^4m_e$, $\mu_1=200m_e$,
and $\mu_2=\sqrt{2}\mu_1$.  The transverse resolution
is $N_\perp=40$.}
\end{figure}
%%%%%%%%%%%%%%%%%%%%%%%%%%%%%%%%%%%%%%%%%%%%%%%%%%%%%%%%%%%%%%
%
%%%%%%%%%%%%%%%%%%%%%%%%%%%%%%%%%%%%%
\begin{table}[ht]
\caption[Same as Table~\ref{tab:aevsK}, but
for the dependence on transverse resolution.]%
{\label{tab:aevsNperp} Same as Table~\ref{tab:aevsK}, but
for the dependence on transverse resolution $N_\perp$.
The longitudinal resolution is $K=30$.}
\begin{tabular}{r|cc|cc}
\hline \hline
      & \multicolumn{2}{c|}{one-photon} & \multicolumn{2}{c|}{with self-energy} \\
\cline{2-5}
$N_\perp$ & $m_0/m_e$ & $2\pi a_e/\alpha$ & $m_0/m_e$ & $2\pi a_e/\alpha$   \\
\hline
10 & 0.98023 & 1.16121 & 0.97918 & --- \\
15 & 0.98297 & 1.15317 & 0.98307 & 0.99071 \\
20 & 0.98268 & 1.15380 & 0.98275 & 0.99776 \\
25 & 0.98256 & 1.15484 & 0.98261 & 1.00130 \\
30 & 0.98248 & 1.15399 & 0.98253 & 1.00348 \\
35 & 0.98244 & 1.15404 & 0.98248 & 1.00501 \\
40 & 0.98241 & 1.15567 & 0.98245 & 1.00612 \\
\hline \hline
\end{tabular}
\end{table}
%%%%%%%%%%%%%%%%%%%%%%%%%%%%%%%%%%%%%%%%%%%%%%%%%%%%%%%%%%%%%%
%
%%%%%%%%%%%%%%%%%%%%%%%%%%%%%%%%%%%%%
\begin{figure}[ht]
\vspace{0.1in}
\centerline{\includegraphics[width=15cm]{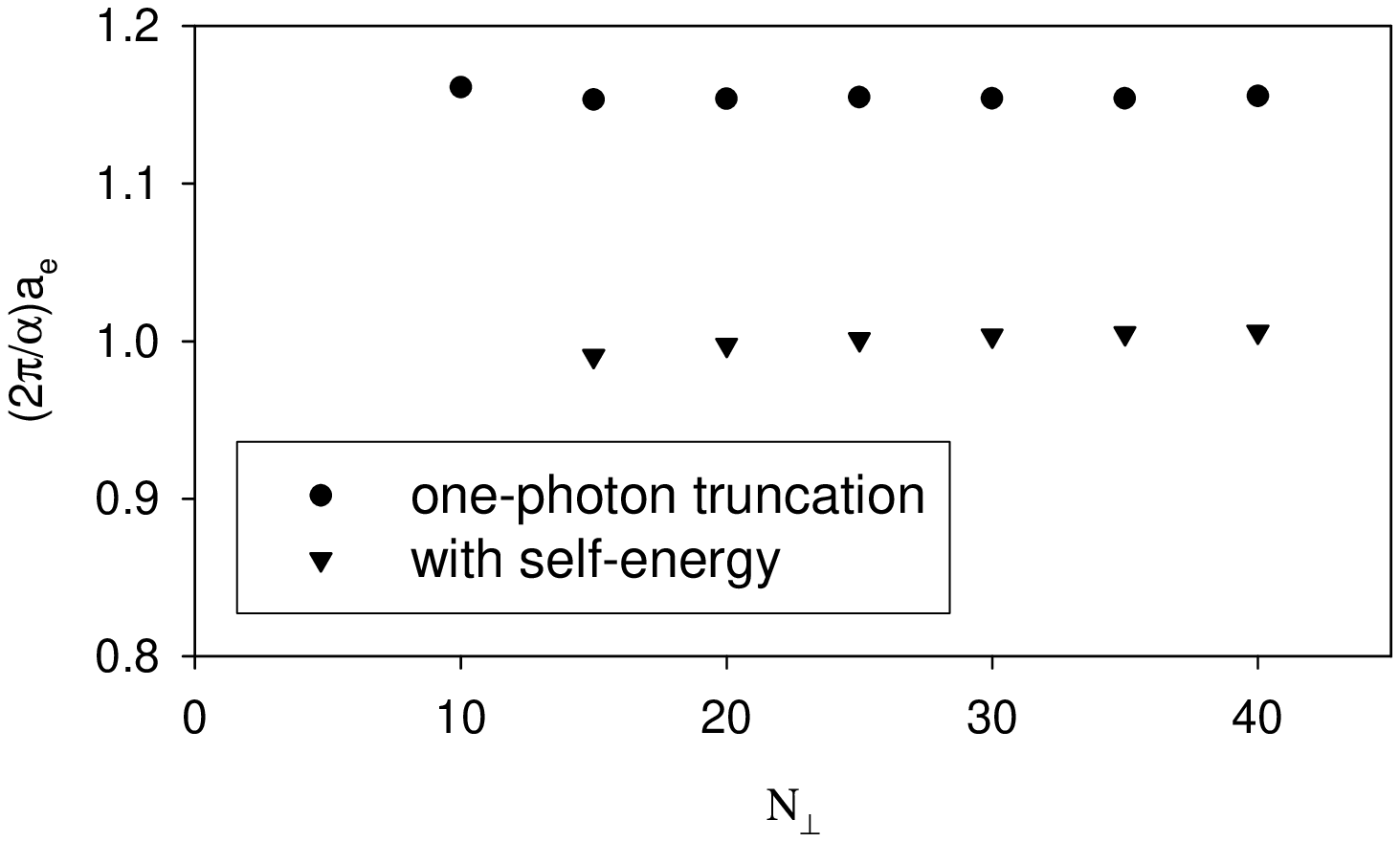}}
\caption[Same as Fig.~\ref{fig:aevsK}, but
for the dependence on transverse resolution.]%
{\label{fig:aevsNperp} 
Same as Fig.~\ref{fig:aevsK}, but
for the dependence on transverse resolution $N_\perp$.
The longitudinal resolution is $K=30$.
}
\end{figure}
%%%%%%%%%%%%%%%%%%%%%%%%%%%%%%%%%%%%%%%%%%%%%%%%%%%%%%%%%%%%%%

As expected, the results for the one-photon truncation,
if computed numerically, converge to better than 1\%
at the same resolution, of $K=20$ and $N_\perp=20$,
as can be seen in Tables~\ref{tab:aevsK} and \ref{tab:aevsNperp}
and Figs.~\ref{fig:aevsK} and \ref{fig:aevsNperp}.
However, the results with the self-energy contribution, shown in
the same tables and figures, require $K\simeq25$ before
nearing convergence.  Although the exact answer is not known
in this case, $K=20$ is clearly insufficient, but $K\gg25$
yields a reasonable result with an error on the order of 1\%.

The two-photon truncation incorporates numerical approximations
to the integrals $\bar I_0$, $\bar I_1$, and $\bar J$ through
the action of the zero-photon kernel $J^{(0)}$, in 
Eq.~(\ref{eq:ReducedEqn}), and approximations to the self-energy
contribution, also in Eq.~(\ref{eq:ReducedEqn}).  Thus, the
minimum resolution for the two-photon calculation would appear
to be approximately $K=25$ and $N_\perp=20$; however, we find
that $K$ must be at least 50.  We extrapolate 
from the one-photon and self-energy calculations 
to estimate an error of 5-10\% for the two-photon truncation.

%%%%%%%%%%%%%%%%%%%%%%%%%%%%%%%%


\begin{thebibliography}{}
%%%%%%%%%%%%%%%%%%%%%%%%%%%%%%%%

%\cite{Kinoshita:2005zr}
\bibitem{Kinoshita}
  T.~Kinoshita and M.~Nio,
  %``Improved alpha**4 term of the electron anomalous magnetic moment,''
  Phys.\ Rev.\  D {\bf 73}, 013003 (2006).
  %[arXiv:hep-ph/0507249].
  %%CITATION = PHRVA,D73,013003;%%

\bibitem{bhm1} S.J. Brodsky, J.R. Hiller, and G. McCartor,
%  Pauli--Villars as a nonperturbative ultraviolet regulator in
%  discretized light-cone quantization,
  Phys.\ Rev.\ D {\bf 58}, 025005 (1998).
  
\bibitem{bhm2} S.J. Brodsky, J.R. Hiller, and G. McCartor,
%  Application of Pauli--Villars regularization and discretized
%  light-cone quantization to a (3+1)-dimensional model,
  Phys.\ Rev.\ D {\bf 60}, 054506 (1999).
  
\bibitem{YukawaDLCQ} S.J. Brodsky, J.R. Hiller, and G. McCartor,
%Application of Pauli--Villars regularization and discretized light-cone
%  quantization to a single-fermion truncation of Yukawa theory},
  Phys.\ Rev.\ D {\bf 64}, 114023 (2001). %, hep-ph/0107038.
  
\bibitem{ExactSolns} S.J. Brodsky, J.R. Hiller, and G. McCartor,
%Exact solutions to Pauli--Villars-regulated field theories},
  Ann.\ Phys.\ {\bf 296}, 406 (2002). %, hep-th/0107246.

\bibitem{YukawaOneBoson} S.J. Brodsky, J.R. Hiller, and G. McCartor,
%The mass renormalization of nonperturbative light-front
%Hamiltonian theory: An illustration using truncated,
%Pauli--Villars-regulated Yukawa interactions,
Ann.\ Phys.\ {\bf 305}, 266 (2003). %, hep-th/0209028.

\bibitem{OnePhotonQED} S.J. Brodsky, V.A. Franke, J.R. Hiller, G. McCartor, 
S.A. Paston, and E.V. Prokhvatilov,
%A nonperturbative calculation of the electron's magnetic moment,
Nucl.\ Phys.\ B {\bf 703}, 333 (2004). %, hep-ph/0406325.

\bibitem{YukawaTwoBoson} S.J. Brodsky, J.R. Hiller, and G. McCartor,
%Two-boson truncation of Pauli--Villars-regulated Yukawa theory,
Ann.\ Phys.\ {\bf 321}, 1240 (2006). %, hep-ph/0508295.

\bibitem{ChiralLimit} S.S. Chabysheva and J.R. Hiller,
 %  Restoration of the chiral limit in Pauli--Villars-regulated light-front QED,
   Phys.\ Rev.\ D {\bf 79}, 114017 (2009).

\bibitem{thesis} S.S. Chabysheva,
   A nonperturbative calculation of the electron's anomalous magnetic moment,
   Ph.D. thesis, Southern Methodist University
   [ProQuest Dissertations \& Theses 3369009, 2009].
   
\bibitem{SecDep} S.S. Chabysheva and J.R. Hiller,
%On the nonperturbative solution of Pauli--Villars regulated light-front QED:
%A comparison of the sector-dependent and standard parameterizations,
Ann.\ Phys.\ {\bf 325}, 2435 (2010).

\bibitem{PauliBrodsky} {H.-C. Pauli and S.J. Brodsky,
   Phys.\ Rev.\ D {\bf 32}, 1993 (1985); {\bf 32}, 2001 (1985).}
   
\bibitem{Varyetal}  J.P.~Vary {\em et al.},
%  Hamiltonian light-front field theory in a basis function approach,
  Phys.\ Rev.\ C {\bf 81}, 035205 (2010).
%  arXiv:0905.1411 [nucl-th].

\bibitem{Lanczos} \label{ref:Lanczos}
   C. Lanczos,
   J. Res.\ Nat.\ Bur.\ Stand.\ {\bf 45}, 255 (1950);
   J.H. Wilkinson,
   {\em The Algebraic Eigenvalue Problem} (Clarendon, Oxford, 1965);
   B.N. Parlett,
   {\em The Symmetric Eigenvalue Problem}
   (Prentice-Hall, Englewood Cliffs, NJ, 1980);
   D.S. Scott, in {\em Sparse Matrices and their Uses}, edited by I.S. Duff
   (Academic Press, London, 1981), p.~139;
   G.H. Golub and C.F. van Loan,
   {\em Matrix Computations} (Johns Hopkins University Press, Baltimore, 1983);
   J. Cullum and R.A. Willoughby,
   in {\em Large-Scale Eigenvalue Problems},
   eds.\ J. Cullum and R.A. Willoughby,
   {\em Math. Stud.} {\bf 127}
   (Elsevier, Amsterdam, 1986), p.~193;
   Y. Saad,
   Comput.\ Phys.\ Commun.\ {\bf 53}, 71 (1989);
   S.K. Kin and A.T. Chronopoulos,
   J. Comp.\ and Appl.\ Math.\ {\bf 42}, 357 (1992).
   
\bibitem{Cullum} J. Cullum and R.A. Willoughby,
   J. Comput.\ Phys. {\bf 44}, 329 (1981);
   {\em Lanczos Algorithms for Large Symmetric Eigenvalue Computations}
   (Birkhauser, Boston, 1985), Vol.\ I and II.

\bibitem{PauliVillars} W. Pauli and F. Villars,
   Rev.\ Mod.\ Phys.\ {\bf 21}, 434 (1949).

\bibitem{Lebed} B. Grinstein, D. O'Connell, and M.B. Wise,
   Phys.\ Rev.\ D {\bf 77}, 025012 (2008);
   C.D. Carone and R.F. Lebed, JHEP {\bf 0901}:043 (2009).
   
\bibitem{Dirac} {P.A.M. Dirac, 
Rev.\ Mod.\ Phys.\ {\bf 21}, 392 (1949).}

\bibitem{DLCQreview} For reviews and additional references, see
   M. Burkardt, Adv.\ Nucl.\ Phys.\ {\bf 23}, 1 (2002);
   S.J. Brodsky, H.-C. Pauli, and S.S. Pinsky, 
   Phys.\ Rep.\ {\bf 301}, 299 (1998). %;
   %J.R. Hiller,
   %Nucl.\ Phys.\ B (Proc.\ Suppl.) {\bf 90}, 170 (2000).

%\cite{Hornbostel:1988fb}
\bibitem{Hornbostel}
  K.~Hornbostel, S.~J.~Brodsky and H.~C.~Pauli,
  %``Light Cone Quantized QCD in (1+1)-Dimensions,''
  Phys.\ Rev.\  D {\bf 41}, 3814 (1990).
  %%CITATION = PHRVA,D41,3814;%%
  
\bibitem{SDLCQ}
Y.~Matsumura, N.~Sakai, and T.~Sakai,
%``Mass spectra of supersymmetric Yang-Mills theories in (1+1)-dimensions,''
Phys.\ Rev.\ D {\bf 52}, 2446 (1995); % [arXiv:hep-th/9504150];
%
%\bibitem{Lunin:1999ib}
O.~Lunin and S.~Pinsky,
%``SDLCQ: Supersymmetric discrete light cone quantization,''
AIP Conf.\ Proc.\  {\bf 494}, 140 (1999); % [arXiv:hep-th/9910222];
%\cite{Hiller:2005vf}
%\bibitem{Hiller:2005vf}
  J.~R.~Hiller, S.~S.~Pinsky, N.~Salwen and U.~Trittmann,
  %``Direct evidence for the Maldacena conjecture for N = (8,8) super
  %Yang-Mills theory in 1+1 dimensions,''
  Phys.\ Lett.\  B {\bf 624}, 105 (2005) %  [arXiv:hep-th/0506225] 
  and references therein.
  %%CITATION = PHLTA,B624,105;%%

\bibitem{EarlyLCQCD} L.C.L. Hollenberg, K. Higashijima, R.C. Warner, and
   B.H.J. McKellar, Prog.\ Th.\ Phys.\ {\bf 87}, 441 (1992).

\bibitem{TammDancoff} {I. Tamm, J. Phys.\ (Moscow) {\bf 9}, 449 (1945);
   S.M. Dancoff, Phys.\ Rev.\ {\bf 78}, 382 (1950).}

\bibitem{SectorDependent} {R.J. Perry, A. Harindranath, and K.G. Wilson,
   Phys.\ Rev.\ Lett.\ {\bf 65}, 2959 (1990);
   R.J. Perry and A. Harindranath, Phys.\ Rev.\ D {\bf 43}, 4051 (1991).}

\bibitem{Wilson} K.G. Wilson, T.S. Walhout, A. Harindranath, 
W.-M. Zhang, R.J. Perry, and St.D. G{\l}azek, 
%Nonperturbative QCD: A Weak coupling treatment on the light front.
Phys.\ Rev.\ D {\bf 49}, 6720 (1994);

\bibitem{RecentPert} {A. Harindranath and R.J. Perry,
   Phys.\ Rev.\ D {\bf 43}, 492 (1991); {\bf 43}, 3580(E) (1991);
   D. Mustaki, S. Pinsky, J. Shigemitsu, and K.G. Wilson,
   {\em ibid}.\ {\bf 43}, 3411 (1991);
   A. Langnau, Ph.D. thesis, SLAC Report 385, 1992;
   A. Langnau and S.J. Brodsky, J. Comput.\ Phys.\ {\bf 109}, 84 (1993);
   R.J. Perry, Phys.\ Lett.\ {\bf B300}, 8 (1993);
   W.-M. Zhang and A. Harindranath,
   Phys.\ Rev.\ D {\bf 48}, 4868 (1993);
   {\bf 48}, 4881 (1993); 
   A. Harindranath and W.-M. Zhang,
   Phys.\ Rev.\ D {\bf 48}, 4903 (1993);
   N.E. Ligterink and B.L.G. Bakker, 
   Phys.\ Rev.\ D {\bf 52}, 5917 (1995); {\bf 52}, 5954 (1995);
   N.C.J. Schoonderwoerd and B.L.G. Bakker, 
   Phys.\ Rev.\ D {\bf 57}, 4965 (1998); {\bf 58}, 025013 (1998).}

\bibitem{Langnau} For a perturbative analysis using light-cone quantization,
   see A. Langnau, Ph.D. thesis, SLAC Report 385, 1992;
   A. Langnau and M. Burkardt, Phys.\ Rev.\ D {\bf 47}, 3452 (1993).

\bibitem{Tang} A.C. Tang, S.J. Brodsky, and H.-C. Pauli,
   Phys.\ Rev.\ D {\bf 44}, 1842 (1991).

\bibitem{Schwinger} {J. Schwinger,
   Phys.\ Rev.\ {\bf 73}, 416 (1948); {\bf 76}, 790 (1949).}
   
\bibitem{SommerfieldPetermann} C.M. Sommerfield,
Phys.\ Rev.\ {\bf 107}, 328 (1957);
A. Petermann, Helv.\ Phys.\ Acta.\ {\bf 30}, 407 (1957).

\bibitem{CQM} See, for example,
%Baryons in a Relativized Quark Model with Chromodynamics.
S. Capstick and N. Isgur, Phys.\ Rev.\ D {\bf 34}, 2809 (1986);
%Mesons in a Relativized Quark Model with Chromodynamics.
S. Godfrey and N. Isgur, Phys.\ Rev.\ D {\bf 32}, 189 (1985).

\bibitem{quarkonia} 
%Quarkonia in Hamiltonian light front QCD.
M.M. Brisudova, R.J. Perry, and K.G. Wilson, 
Phys.\ Rev.\ Lett.\ {\bf 78}, 1227 (1997).

\bibitem{AdSQCD}   S.~J.~Brodsky and G.~F.~de~Teramond,
  %``Hadronic spectra and light-front wavefunctions in holographic QCD,''
  Phys.\ Rev.\ Lett.\  {\bf 96}, 201601 (2006). %   [arXiv:hep-ph/0602252].
  %%CITATION = PRLTA,96,201601;%%

\bibitem{Glazek}  S.~D.~G{\l}azek and J.~Mlynik,
  %``Boost-invariant Hamiltonian approach to heavy quarkonia,''
  Phys.\ Rev.\ D {\bf 74}, 105015 (2006), %, hep-th/0606235;
  %%CITATION = HEP-TH 0606235;
  S.~D.~G{\l}azek,
  %``Harmonic oscillator force between heavy quarks,''
  Phys.\ Rev.\ D {\bf 69}, 065002 (2004), %,  hep-th/0307064;
  %%CITATION = HEP-TH 0307064;%%
  S.~D.~G{\l}azek and J.~Mlynik,
  Phys.\ Rev.\ D {\bf 67}, 045001 (2003);
  St.D.~G{\l}azek and M.~Wieckowski,
  Phys.\ Rev.\ D {\bf 66}, 016001 (2002).

\bibitem{hb} J.R. Hiller and S.J. Brodsky,
%  Nonperturbative renormalization and the electron's 
%  anomalous moment in large-$\alpha$ QED,
  Phys.\ Rev.\ D {\bf 59}, 016006 (1998).
  
\bibitem{Karmanov}
  V.~A.~Karmanov, J.~F.~Mathiot, and A.~V.~Smirnov,
  Phys.\ Rev.\ D {\bf 77}, 085028 (2008).
  
\bibitem{lattice} For reviews, see M. Creutz, L. Jacobs and C. Rebbi,
   Phys.\ Rep.\ {\bf 95}, 201 (1983);
   J.B. Kogut, Rev.\ Mod.\ Phys.\ {\bf 55}, 775 (1983);
   I. Montvay, {\em ibid}.\ {\bf 59}, 263 (1987);
   A.S. Kronfeld and P.B. Mackenzie,
   Ann.\ Rev.\ Nucl.\ Part.\ Sci.\ {\bf 43}, 793 (1993);
   J.W. Negele, Nucl.\ Phys.\ {\bf A553}, 47c (1993);
   K.G.~Wilson,
%   ``The origins of lattice gauge theory,''
   Nucl.\ Phys.\ B (Proc.\ Suppl.) {\bf 140}, 3 (2005);
   J.M. Zanotti, 
   %Investigations of hadron structure on the lattice,
   PoS {\bf LAT2008}, 007 (2008).
   For recent discussions of meson properties and charm physics, see for example
   C. McNeile and C. Michael [UKQCD Collaboration], 
   Phys.\ Rev.\ D {\bf 74}, 014508 (2006);
   I.~Allison {\em et al.}  [HPQCD Collaboration],
  %``High-Precision Charm-Quark Mass from Current-Current Correlators in Lattice
  %and Continuum QCD,''
   Phys.\ Rev.\  D {\bf 78}, 054513 (2008).  %  [arXiv:0805.2999 [hep-lat]].
   
\bibitem{TransLattice} M.~Burkardt and S.~Dalley,
   Prog.\ Part.\ Nucl.\ Phys. \ {\bf 48}, 317 (2002) and references therein; 
%   M.~Burkardt and H.~El-Khozondar,
%  %``Wilson fermions on a transverse lattice,''
%   Phys.\ Rev.\  D {\bf 60}, 054504 (1999);  %  [arXiv:hep-ph/9805495].
%   M.~Burkardt and S.~K.~Seal,
%  %``A study of heavy light mesons on the transverse lattice,''
%   Phys.\ Rev.\  D {\bf 64}, 111501 (2001); %  [arXiv:hep-ph/0105109].
%  %``A study of light mesons on the transverse lattice,''
%   Phys.\ Rev.\  D {\bf 65}, 034501 (2002); %  [arXiv:hep-ph/0102245].
   S.~Dalley and B.~van~de~Sande, 
   Phys.\ Rev.\ D {\bf 67}, 114507 (2003);
   D. Chakrabarti, A.K. De, and A. Harindranath,
   Phys.\ Rev.\ D {\bf 67}, 076004 (2003);
   M. Harada and S. Pinsky, 
   Phys.\ Lett.\ B {\bf 567}, 277 (2003);
   S.~Dalley and B.~van de Sande,
  %``Finite temperature gauge theory from the transverse lattice,''
   Phys.\ Rev.\ Lett.\  {\bf 95}, 162001 (2005); %  [arXiv:hep-ph/0409114].
   J.~Bratt, S.~Dalley, B.~van de Sande, and E.~M.~Watson,
  %``Small-x behaviour of lightcone wavefunctions in transverse lattice  gauge
  %theory,''
   Phys.\ Rev.\  D {\bf 70}, 114502 (2004). %  [arXiv:hep-ph/0410188].
   For work on a complete light-cone lattice, see
   C. Destri and H.J. de Vega,
   Nucl.\ Phys.\ {\bf B290}, 363 (1987);
   D. Mustaki, Phys.\ Rev.\ D {\bf 38}, 1260 (1988).

\bibitem{SchwingerDyson} C.D. Roberts and A.G. Williams,
   Prog.\ Part.\ Nucl.\ Phys.\ {\bf 33}, 477 (1994);
   P. Maris and C.D. Roberts, Int.\ J. Mod.\ Phys.\ {\bf E12}, 297 (2003);
   P.C. Tandy, Nucl.\ Phys.\ B (Proc.\ Suppl.) {\bf 141}, 9 (2005).
   
\bibitem{BRS} S.J. Brodsky, R. Roskies, and R. Suaya,
   Phys.\ Rev.\ D {\bf 8}, 4574 (1973).

\bibitem{BrodskyDrell} S.J. Brodsky and S.D. Drell,
Phys.\ Rev.\ D {\bf 22}, 2236 (1980).

\bibitem{BurdenFaires} R.L. Burden and J.D. Faires,
  {\em Numerical Analysis}, 3rd ed., (Prindle, Weber \& Schmidt, Boston, 1985).
%W.H. Press, S.A. Teukolsky, W.T. Vetterling, and B.P. Flannery,
%{\em Numerical Recipes in Fortran 77: The Art of Scientific Computing}, 2nd ed.,
%(Cambridge University Press, Cambridge, 1992).

\bibitem{Positronium} 
   D. Klabucar and H.-C. Pauli, Z. Phys.\ C {\bf 47}, 141 (1990);
   M. Kalu\v{z}a and H.-C. Pauli, 
   Phys.\ Rev.\ D {\bf 45}, 2968 (1992);
   M. Krautg\"{a}rtner, H.C. Pauli, and F. W\"{o}lz,
   {\em ibid}.\ {\bf 45}, 3755 (1992);
   U. Trittmann and H.-C. Pauli, hep-th/9704215; hep-th/9705021;
   U. Trittmann, hep-th/9705072; hep-th/9706055.

\bibitem{Paston}
S.A.~Paston and V.A.~Franke,
%``Comparison of quantum field perturbation theory for the light front
%with the theory in lorentz coordinates,'' 
Theor.\ Math.\ Phys.\  \textbf{112}, 1117 (1997) 
[Teor.\ Mat.\ Fiz.\  \textbf{112}, 399 (1997)];
%[arXiv:hep-th/9901110]; 
%%CITATION = HEP-TH 9901110;%%
%\cite{Paston:2000fq}
S.A.~Paston, V.A.~Franke, and E.V.~Prokhvatilov,
%``Constructing the light-front QCD Hamiltonian,''
Theor.\ Math.\ Phys.\  \textbf{120}, 1164 (1999)
[Teor.\ Mat.\ Fiz.\  \textbf{120}, 417 (1999)].  %;
%[arXiv:hep-th/0002062].
%%CITATION = HEP-TH 0002062;%%
 
\bibitem{DalleyMcCartor}
  S.~Dalley and G.~McCartor,
  %``Spontaneously broken quark helicity symmetry,''
  Ann.\ Phys.\  {\bf 321}, 402 (2006).  %,
 % hep-ph/0406287.

\end{thebibliography}
\end{document}